\documentclass[twocolumn,traditabstract]{aa}
\usepackage[breaklinks, colorlinks, citecolor=blue]{hyperref}
\usepackage{graphicx,amsmath}
\usepackage{epstopdf}
\usepackage{epsf,color}
\usepackage{txfonts}
\usepackage[nonamebreak]{natbib}
\usepackage{mathrsfs}
\usepackage{amsmath}
\bibpunct{(}{)}{;}{a}{}{,} 

\def\setsymbol#1#2{\expandafter\def\csname #1\endcsname{#2}}
\def\getsymbol#1{\csname #1\endcsname}

\def\Planck{{\it Planck\/}}

\newbox\tablebox    \newdimen\tablewidth
\def\leaderfil{\leaders\hbox to 5pt{\hss.\hss}\hfil}
\def\endPlancktablewide{\tablewidth=\textwidth 
    $$\hss\copy\tablebox\hss$$
    \vskip-\lastskip\vskip -2pt}
\def\tablenote#1 #2\par{\begingroup \parindent=0.8em
    \abovedisplayshortskip=0pt\belowdisplayshortskip=0pt
    \noindent
    $$\hss\vbox{\hsize\tablewidth \hangindent=\parindent \hangafter=1 \noindent
    \hbox to \parindent{\sup{\rm #1}\hss}\strut#2\strut\par}\hss$$
    \endgroup}
\def\doubleline{\vskip 3pt\hrule \vskip 1.5pt \hrule \vskip 5pt}

\def\L2{\ifmmode L_2\else $L_2$\fi}

\def\DeltaT{\ifmmode \Delta T\else $\Delta T$\fi}
\def\deltat{\ifmmode \Delta t\else $\Delta t$\fi}
\def\fknee{\ifmmode f_{\rm knee}\else $f_{\rm knee}$\fi}
\def\Fmax{\ifmmode F_{\rm max}\else $F_{\rm max}$\fi}
\def\solar{\ifmmode{\rm M}_{\mathord\odot}\else${\rm M}_{\mathord\odot}$\fi}

\def\inv{\ifmmode^{-1}\else$^{-1}$\fi}
\def\mo{\ifmmode^{-1}\else$^{-1}$\fi}
\def\sup#1{\ifmmode ^{\rm #1}\else $^{\rm #1}$\fi}
\def\expo#1{\ifmmode \times 10^{#1}\else $\times 10^{#1}$\fi}
\def\,{\thinspace}
\def\lsim{\mathrel{\raise .4ex\hbox{\rlap{$<$}\lower 1.2ex\hbox{$\sim$}}}}
\def\gsim{\mathrel{\raise .4ex\hbox{\rlap{$>$}\lower 1.2ex\hbox{$\sim$}}}}

\def\simprop{\mathrel{\raise .4ex\hbox{\rlap{$\propto$}\lower 1.2ex\hbox{$\sim$}}}}
\def\deg{\ifmmode^\circ\else$^\circ$\fi}
\def\pdeg{\ifmmode $\setbox0=\hbox{$^{\circ}$}\rlap{\hskip.11\wd0 .}$^{\circ}
          \else \setbox0=\hbox{$^{\circ}$}\rlap{\hskip.11\wd0 .}$^{\circ}$\fi}
\def\arcs{\ifmmode {^{\scriptstyle\prime\prime}}
          \else $^{\scriptstyle\prime\prime}$\fi}
\def\arcm{\ifmmode {^{\scriptstyle\prime}}
          \else $^{\scriptstyle\prime}$\fi}
\newdimen\sa  \newdimen\sb
\def\parcs{\sa=.07em \sb=.03em
     \ifmmode \hbox{\rlap{.}}^{\scriptstyle\prime\kern -\sb\prime}\hbox{\kern -\sa}
     \else \rlap{.}$^{\scriptstyle\prime\kern -\sb\prime}$\kern -\sa\fi}
\def\parcm{\sa=.08em \sb=.03em
     \ifmmode \hbox{\rlap{.}\kern\sa}^{\scriptstyle\prime}\hbox{\kern-\sb}
     \else \rlap{.}\kern\sa$^{\scriptstyle\prime}$\kern-\sb\fi}
\def\ra[#1 #2 #3.#4]{#1\sup{h}#2\sup{m}#3\sup{s}\llap.#4}
\def\dec[#1 #2 #3.#4]{#1\deg#2\arcm#3\arcs\llap.#4}
\def\deco[#1 #2 #3]{#1\deg#2\arcm#3\arcs}
\def\rra[#1 #2]{#1\sup{h}#2\sup{m}}

\def\dots{\relax\ifmmode \ldots\else $\ldots$\fi}
\def\WHzsr{\ifmmode $W\,Hz\mo\,sr\mo$\else W\,Hz\mo\,sr\mo\fi}
\def\mHz{\ifmmode $\,mHz$\else \,mHz\fi}
\def\GHz{\ifmmode $\,GHz$\else \,GHz\fi}
\def\mKs{\ifmmode $\,mK\,s$^{1/2}\else \,mK\,s$^{1/2}$\fi}
\def\muKs{\ifmmode \,\mu$K\,s$^{1/2}\else \,$\mu$K\,s$^{1/2}$\fi}
\def\muKRJs{\ifmmode \,\mu$K$_{\rm RJ}$\,s$^{1/2}\else \,$\mu$K$_{\rm RJ}$\,s$^{1/2}$\fi}
\def\muKHz{\ifmmode \,\mu$K\,Hz$^{-1/2}\else \,$\mu$K\,Hz$^{-1/2}$\fi}
\def\MJysr{\ifmmode \,$MJy\,sr\mo$\else \,MJy\,sr\mo\fi}
\def\MJysrmK{\ifmmode \,$MJy\,sr\mo$\,mK$_{\rm CMB}\mo\else \,MJy\,sr\mo\,mK$_{\rm CMB}\mo$\fi}
\def\microns{\ifmmode \,\mu$m$\else \,$\mu$m\fi}

\def\muK{\ifmmode \,\mu$K$\else \,$\mu$\hbox{K}\fi}
\def\microK{\ifmmode \,\mu$K$\else \,$\mu$\hbox{K}\fi}
\def\muW{\ifmmode \,\mu$W$\else \,$\mu$\hbox{W}\fi}
\def\kms{\ifmmode $\,km\,s$^{-1}\else \,km\,s$^{-1}$\fi}
\def\kmsMpc{\ifmmode $\,\kms\,Mpc\mo$\else \,\kms\,Mpc\mo\fi}
\setsymbol{LFI:center:frequency:70GHz:units}{70.3\,GHz}
\setsymbol{LFI:center:frequency:44GHz:units}{44.1\,GHz}
\setsymbol{LFI:center:frequency:30GHz:units}{28.5\,GHz}

\setsymbol{LFI:center:frequency:70GHz}{70.3}
\setsymbol{LFI:center:frequency:44GHz}{44.1}
\setsymbol{LFI:center:frequency:30GHz}{28.5}

\setsymbol{LFI:center:frequency:LFI18:Rad:M:units}{71.7\GHz}
\setsymbol{LFI:center:frequency:LFI19:Rad:M:units}{67.5\GHz}
\setsymbol{LFI:center:frequency:LFI20:Rad:M:units}{69.2\GHz}
\setsymbol{LFI:center:frequency:LFI21:Rad:M:units}{70.4\GHz}
\setsymbol{LFI:center:frequency:LFI22:Rad:M:units}{71.5\GHz}
\setsymbol{LFI:center:frequency:LFI23:Rad:M:units}{70.8\GHz}
\setsymbol{LFI:center:frequency:LFI24:Rad:M:units}{44.4\GHz}
\setsymbol{LFI:center:frequency:LFI25:Rad:M:units}{44.0\GHz}
\setsymbol{LFI:center:frequency:LFI26:Rad:M:units}{43.9\GHz}
\setsymbol{LFI:center:frequency:LFI27:Rad:M:units}{28.3\GHz}
\setsymbol{LFI:center:frequency:LFI28:Rad:M:units}{28.8\GHz}
\setsymbol{LFI:center:frequency:LFI18:Rad:S:units}{70.1\GHz}
\setsymbol{LFI:center:frequency:LFI19:Rad:S:units}{69.6\GHz}
\setsymbol{LFI:center:frequency:LFI20:Rad:S:units}{69.5\GHz}
\setsymbol{LFI:center:frequency:LFI21:Rad:S:units}{69.5\GHz}
\setsymbol{LFI:center:frequency:LFI22:Rad:S:units}{72.8\GHz}
\setsymbol{LFI:center:frequency:LFI23:Rad:S:units}{71.3\GHz}
\setsymbol{LFI:center:frequency:LFI24:Rad:S:units}{44.1\GHz}
\setsymbol{LFI:center:frequency:LFI25:Rad:S:units}{44.1\GHz}
\setsymbol{LFI:center:frequency:LFI26:Rad:S:units}{44.1\GHz}
\setsymbol{LFI:center:frequency:LFI27:Rad:S:units}{28.5\GHz}
\setsymbol{LFI:center:frequency:LFI28:Rad:S:units}{28.2\GHz}

\setsymbol{LFI:center:frequency:LFI18:Rad:M}{71.7}
\setsymbol{LFI:center:frequency:LFI19:Rad:M}{67.5}
\setsymbol{LFI:center:frequency:LFI20:Rad:M}{69.2}
\setsymbol{LFI:center:frequency:LFI21:Rad:M}{70.4}
\setsymbol{LFI:center:frequency:LFI22:Rad:M}{71.5}
\setsymbol{LFI:center:frequency:LFI23:Rad:M}{70.8}
\setsymbol{LFI:center:frequency:LFI24:Rad:M}{44.4}
\setsymbol{LFI:center:frequency:LFI25:Rad:M}{44.0}
\setsymbol{LFI:center:frequency:LFI26:Rad:M}{43.9}
\setsymbol{LFI:center:frequency:LFI27:Rad:M}{28.3}
\setsymbol{LFI:center:frequency:LFI28:Rad:M}{28.8}
\setsymbol{LFI:center:frequency:LFI18:Rad:S}{70.1}
\setsymbol{LFI:center:frequency:LFI19:Rad:S}{69.6}
\setsymbol{LFI:center:frequency:LFI20:Rad:S}{69.5}
\setsymbol{LFI:center:frequency:LFI21:Rad:S}{69.5}
\setsymbol{LFI:center:frequency:LFI22:Rad:S}{72.8}
\setsymbol{LFI:center:frequency:LFI23:Rad:S}{71.3}
\setsymbol{LFI:center:frequency:LFI24:Rad:S}{44.1}
\setsymbol{LFI:center:frequency:LFI25:Rad:S}{44.1}
\setsymbol{LFI:center:frequency:LFI26:Rad:S}{44.1}
\setsymbol{LFI:center:frequency:LFI27:Rad:S}{28.5}
\setsymbol{LFI:center:frequency:LFI28:Rad:S}{28.2}

\setsymbol{LFI:white:noise:sensitivity:70GHz:units}{134.7\muKs}
\setsymbol{LFI:white:noise:sensitivity:44GHz:units}{164.7\muKs}
\setsymbol{LFI:white:noise:sensitivity:30GHz:units}{143.4\muKs}

\setsymbol{LFI:white:noise:sensitivity:70GHz}{134.7}
\setsymbol{LFI:white:noise:sensitivity:44GHz}{164.7}
\setsymbol{LFI:white:noise:sensitivity:30GHz}{143.4}

\setsymbol{LFI:white:noise:sensitivity:LFI18:Rad:M:units}{512.0\muKs}
\setsymbol{LFI:white:noise:sensitivity:LFI19:Rad:M:units}{581.4\muKs}
\setsymbol{LFI:white:noise:sensitivity:LFI20:Rad:M:units}{590.8\muKs}
\setsymbol{LFI:white:noise:sensitivity:LFI21:Rad:M:units}{455.2\muKs}
\setsymbol{LFI:white:noise:sensitivity:LFI22:Rad:M:units}{492.0\muKs}
\setsymbol{LFI:white:noise:sensitivity:LFI23:Rad:M:units}{507.7\muKs}
\setsymbol{LFI:white:noise:sensitivity:LFI24:Rad:M:units}{462.2\muKs}
\setsymbol{LFI:white:noise:sensitivity:LFI25:Rad:M:units}{413.6\muKs}
\setsymbol{LFI:white:noise:sensitivity:LFI26:Rad:M:units}{478.6\muKs}
\setsymbol{LFI:white:noise:sensitivity:LFI27:Rad:M:units}{277.7\muKs}
\setsymbol{LFI:white:noise:sensitivity:LFI28:Rad:M:units}{312.3\muKs}
\setsymbol{LFI:white:noise:sensitivity:LFI18:Rad:S:units}{465.7\muKs}
\setsymbol{LFI:white:noise:sensitivity:LFI19:Rad:S:units}{555.6\muKs}
\setsymbol{LFI:white:noise:sensitivity:LFI20:Rad:S:units}{623.2\muKs}
\setsymbol{LFI:white:noise:sensitivity:LFI21:Rad:S:units}{564.1\muKs}
\setsymbol{LFI:white:noise:sensitivity:LFI22:Rad:S:units}{534.4\muKs}
\setsymbol{LFI:white:noise:sensitivity:LFI23:Rad:S:units}{542.4\muKs}
\setsymbol{LFI:white:noise:sensitivity:LFI24:Rad:S:units}{399.2\muKs}
\setsymbol{LFI:white:noise:sensitivity:LFI25:Rad:S:units}{392.6\muKs}
\setsymbol{LFI:white:noise:sensitivity:LFI26:Rad:S:units}{418.6\muKs}
\setsymbol{LFI:white:noise:sensitivity:LFI27:Rad:S:units}{302.9\muKs}
\setsymbol{LFI:white:noise:sensitivity:LFI28:Rad:S:units}{285.3\muKs}

\setsymbol{LFI:white:noise:sensitivity:LFI18:Rad:M}{512.0}
\setsymbol{LFI:white:noise:sensitivity:LFI19:Rad:M}{581.4}
\setsymbol{LFI:white:noise:sensitivity:LFI20:Rad:M}{590.8}
\setsymbol{LFI:white:noise:sensitivity:LFI21:Rad:M}{455.2}
\setsymbol{LFI:white:noise:sensitivity:LFI22:Rad:M}{492.0}
\setsymbol{LFI:white:noise:sensitivity:LFI23:Rad:M}{507.7}
\setsymbol{LFI:white:noise:sensitivity:LFI24:Rad:M}{462.2}
\setsymbol{LFI:white:noise:sensitivity:LFI25:Rad:M}{413.6}
\setsymbol{LFI:white:noise:sensitivity:LFI26:Rad:M}{478.6}
\setsymbol{LFI:white:noise:sensitivity:LFI27:Rad:M}{277.7}
\setsymbol{LFI:white:noise:sensitivity:LFI28:Rad:M}{312.3}
\setsymbol{LFI:white:noise:sensitivity:LFI18:Rad:S}{465.7}
\setsymbol{LFI:white:noise:sensitivity:LFI19:Rad:S}{555.6}
\setsymbol{LFI:white:noise:sensitivity:LFI20:Rad:S}{623.2}
\setsymbol{LFI:white:noise:sensitivity:LFI21:Rad:S}{564.1}
\setsymbol{LFI:white:noise:sensitivity:LFI22:Rad:S}{534.4}
\setsymbol{LFI:white:noise:sensitivity:LFI23:Rad:S}{542.4}
\setsymbol{LFI:white:noise:sensitivity:LFI24:Rad:S}{399.2}
\setsymbol{LFI:white:noise:sensitivity:LFI25:Rad:S}{392.6}
\setsymbol{LFI:white:noise:sensitivity:LFI26:Rad:S}{418.6}
\setsymbol{LFI:white:noise:sensitivity:LFI27:Rad:S}{302.9}
\setsymbol{LFI:white:noise:sensitivity:LFI28:Rad:S}{285.3}

\setsymbol{LFI:knee:frequency:70GHz:units}{29.5\mHz}
\setsymbol{LFI:knee:frequency:44GHz:units}{56.2\mHz}
\setsymbol{LFI:knee:frequency:30GHz:units}{113.7\mHz}

\setsymbol{LFI:knee:frequency:70GHz}{29.5}
\setsymbol{LFI:knee:frequency:44GHz}{56.2}
\setsymbol{LFI:knee:frequency:30GHz}{113.7}

\setsymbol{LFI:knee:frequency:LFI18:Rad:M:units}{16.3\mHz}
\setsymbol{LFI:knee:frequency:LFI19:Rad:M:units}{15.1\mHz}
\setsymbol{LFI:knee:frequency:LFI20:Rad:M:units}{18.7\mHz}
\setsymbol{LFI:knee:frequency:LFI21:Rad:M:units}{37.2\mHz}
\setsymbol{LFI:knee:frequency:LFI22:Rad:M:units}{12.7\mHz}
\setsymbol{LFI:knee:frequency:LFI23:Rad:M:units}{34.6\mHz}
\setsymbol{LFI:knee:frequency:LFI24:Rad:M:units}{46.2\mHz}
\setsymbol{LFI:knee:frequency:LFI25:Rad:M:units}{24.9\mHz}
\setsymbol{LFI:knee:frequency:LFI26:Rad:M:units}{67.6\mHz}
\setsymbol{LFI:knee:frequency:LFI27:Rad:M:units}{187.4\mHz}
\setsymbol{LFI:knee:frequency:LFI28:Rad:M:units}{122.2\mHz}
\setsymbol{LFI:knee:frequency:LFI18:Rad:S:units}{17.7\mHz}
\setsymbol{LFI:knee:frequency:LFI19:Rad:S:units}{22.0\mHz}
\setsymbol{LFI:knee:frequency:LFI20:Rad:S:units}{8.7\mHz}
\setsymbol{LFI:knee:frequency:LFI21:Rad:S:units}{25.9\mHz}
\setsymbol{LFI:knee:frequency:LFI22:Rad:S:units}{15.8\mHz}
\setsymbol{LFI:knee:frequency:LFI23:Rad:S:units}{129.8\mHz}
\setsymbol{LFI:knee:frequency:LFI24:Rad:S:units}{100.9\mHz}
\setsymbol{LFI:knee:frequency:LFI25:Rad:S:units}{38.9\mHz}
\setsymbol{LFI:knee:frequency:LFI26:Rad:S:units}{58.9\mHz}
\setsymbol{LFI:knee:frequency:LFI27:Rad:S:units}{104.4\mHz}
\setsymbol{LFI:knee:frequency:LFI28:Rad:S:units}{40.7\mHz}

\setsymbol{LFI:knee:frequency:LFI18:Rad:M}{16.3}
\setsymbol{LFI:knee:frequency:LFI19:Rad:M}{15.1}
\setsymbol{LFI:knee:frequency:LFI20:Rad:M}{18.7}
\setsymbol{LFI:knee:frequency:LFI21:Rad:M}{37.2}
\setsymbol{LFI:knee:frequency:LFI22:Rad:M}{12.7}
\setsymbol{LFI:knee:frequency:LFI23:Rad:M}{34.6}
\setsymbol{LFI:knee:frequency:LFI24:Rad:M}{46.2}
\setsymbol{LFI:knee:frequency:LFI25:Rad:M}{24.9}
\setsymbol{LFI:knee:frequency:LFI26:Rad:M}{67.6}
\setsymbol{LFI:knee:frequency:LFI27:Rad:M}{187.4}
\setsymbol{LFI:knee:frequency:LFI28:Rad:M}{122.2}
\setsymbol{LFI:knee:frequency:LFI18:Rad:S}{17.7}
\setsymbol{LFI:knee:frequency:LFI19:Rad:S}{22.0}
\setsymbol{LFI:knee:frequency:LFI20:Rad:S}{8.7}
\setsymbol{LFI:knee:frequency:LFI21:Rad:S}{25.9}
\setsymbol{LFI:knee:frequency:LFI22:Rad:S}{15.8}
\setsymbol{LFI:knee:frequency:LFI23:Rad:S}{129.8}
\setsymbol{LFI:knee:frequency:LFI24:Rad:S}{100.9}
\setsymbol{LFI:knee:frequency:LFI25:Rad:S}{38.9}
\setsymbol{LFI:knee:frequency:LFI26:Rad:S}{58.9}
\setsymbol{LFI:knee:frequency:LFI27:Rad:S}{104.4}
\setsymbol{LFI:knee:frequency:LFI28:Rad:S}{40.7}

\setsymbol{LFI:slope:70GHz:units}{$-1.03$\mHz}
\setsymbol{LFI:slope:44GHz:units}{$-0.89$\mHz}
\setsymbol{LFI:slope:30GHz:units}{$-0.87$\mHz}

\setsymbol{LFI:slope:70GHz}{$-1.03$}
\setsymbol{LFI:slope:44GHz}{$-0.89$}
\setsymbol{LFI:slope:30GHz}{$-0.87$}

\setsymbol{LFI:slope:LFI18:Rad:M:units}{$-1.04$\mHz}
\setsymbol{LFI:slope:LFI19:Rad:M:units}{$-1.09$\mHz}
\setsymbol{LFI:slope:LFI20:Rad:M:units}{$-0.69$\mHz}
\setsymbol{LFI:slope:LFI21:Rad:M:units}{$-1.56$\mHz}
\setsymbol{LFI:slope:LFI22:Rad:M:units}{$-1.01$\mHz}
\setsymbol{LFI:slope:LFI23:Rad:M:units}{$-0.96$\mHz}
\setsymbol{LFI:slope:LFI24:Rad:M:units}{$-0.83$\mHz}
\setsymbol{LFI:slope:LFI25:Rad:M:units}{$-0.91$\mHz}
\setsymbol{LFI:slope:LFI26:Rad:M:units}{$-0.95$\mHz}
\setsymbol{LFI:slope:LFI27:Rad:M:units}{$-0.87$\mHz}
\setsymbol{LFI:slope:LFI28:Rad:M:units}{$-0.88$\mHz}
\setsymbol{LFI:slope:LFI18:Rad:S:units}{$-1.15$\mHz}
\setsymbol{LFI:slope:LFI19:Rad:S:units}{$-1.00$\mHz}
\setsymbol{LFI:slope:LFI20:Rad:S:units}{$-0.95$\mHz}
\setsymbol{LFI:slope:LFI21:Rad:S:units}{$-0.92$\mHz}
\setsymbol{LFI:slope:LFI22:Rad:S:units}{$-1.01$\mHz}
\setsymbol{LFI:slope:LFI23:Rad:S:units}{$-0.95$\mHz}
\setsymbol{LFI:slope:LFI24:Rad:S:units}{$-0.73$\mHz}
\setsymbol{LFI:slope:LFI25:Rad:S:units}{$-1.16$\mHz}
\setsymbol{LFI:slope:LFI26:Rad:S:units}{$-0.79$\mHz}
\setsymbol{LFI:slope:LFI27:Rad:S:units}{$-0.82$\mHz}
\setsymbol{LFI:slope:LFI28:Rad:S:units}{$-0.91$\mHz}

\setsymbol{LFI:slope:LFI18:Rad:M}{$-1.04$}
\setsymbol{LFI:slope:LFI19:Rad:M}{$-1.09$}
\setsymbol{LFI:slope:LFI20:Rad:M}{$-0.69$}
\setsymbol{LFI:slope:LFI21:Rad:M}{$-1.56$}
\setsymbol{LFI:slope:LFI22:Rad:M}{$-1.01$}
\setsymbol{LFI:slope:LFI23:Rad:M}{$-0.96$}
\setsymbol{LFI:slope:LFI24:Rad:M}{$-0.83$}
\setsymbol{LFI:slope:LFI25:Rad:M}{$-0.91$}
\setsymbol{LFI:slope:LFI26:Rad:M}{$-0.95$}
\setsymbol{LFI:slope:LFI27:Rad:M}{$-0.87$}
\setsymbol{LFI:slope:LFI28:Rad:M}{$-0.88$}
\setsymbol{LFI:slope:LFI18:Rad:S}{$-1.15$}
\setsymbol{LFI:slope:LFI19:Rad:S}{$-1.00$}
\setsymbol{LFI:slope:LFI20:Rad:S}{$-0.95$}
\setsymbol{LFI:slope:LFI21:Rad:S}{$-0.92$}
\setsymbol{LFI:slope:LFI22:Rad:S}{$-1.01$}
\setsymbol{LFI:slope:LFI23:Rad:S}{$-0.95$}
\setsymbol{LFI:slope:LFI24:Rad:S}{$-0.73$}
\setsymbol{LFI:slope:LFI25:Rad:S}{$-1.16$}
\setsymbol{LFI:slope:LFI26:Rad:S}{$-0.79$}
\setsymbol{LFI:slope:LFI27:Rad:S}{$-0.82$}
\setsymbol{LFI:slope:LFI28:Rad:S}{$-0.91$}

\setsymbol{LFI:FWHM:70GHz:units}{13\parcm01}
\setsymbol{LFI:FWHM:44GHz:units}{27\parcm92}
\setsymbol{LFI:FWHM:30GHz:units}{32\parcm65}

\setsymbol{LFI:FWHM:70GHz}{13.01}
\setsymbol{LFI:FWHM:44GHz}{27.92}
\setsymbol{LFI:FWHM:30GHz}{32.65}

\setsymbol{LFI:FWHM:LFI18:units}{13\parcm39}
\setsymbol{LFI:FWHM:LFI19:units}{13\parcm01}
\setsymbol{LFI:FWHM:LFI20:units}{12\parcm75}
\setsymbol{LFI:FWHM:LFI21:units}{12\parcm74}
\setsymbol{LFI:FWHM:LFI22:units}{12\parcm87}
\setsymbol{LFI:FWHM:LFI23:units}{13\parcm27}
\setsymbol{LFI:FWHM:LFI24:units}{22\parcm98}
\setsymbol{LFI:FWHM:LFI25:units}{30\parcm46}
\setsymbol{LFI:FWHM:LFI26:units}{30\parcm31}
\setsymbol{LFI:FWHM:LFI27:units}{32\parcm65}
\setsymbol{LFI:FWHM:LFI28:units}{32\parcm66}

\setsymbol{LFI:FWHM:LFI18}{13.39}
\setsymbol{LFI:FWHM:LFI19}{13.01}
\setsymbol{LFI:FWHM:LFI20}{12.75}
\setsymbol{LFI:FWHM:LFI21}{12.74}
\setsymbol{LFI:FWHM:LFI22}{12.87}
\setsymbol{LFI:FWHM:LFI23}{13.27}
\setsymbol{LFI:FWHM:LFI24}{22.98}
\setsymbol{LFI:FWHM:LFI25}{30.46}
\setsymbol{LFI:FWHM:LFI26}{30.31}
\setsymbol{LFI:FWHM:LFI27}{32.65}
\setsymbol{LFI:FWHM:LFI28}{32.66}

\setsymbol{LFI:FWHM:uncertainty:LFI18:units}{0.170\arcm}
\setsymbol{LFI:FWHM:uncertainty:LFI19:units}{0.174\arcm}
\setsymbol{LFI:FWHM:uncertainty:LFI20:units}{0.170\arcm}
\setsymbol{LFI:FWHM:uncertainty:LFI21:units}{0.156\arcm}
\setsymbol{LFI:FWHM:uncertainty:LFI22:units}{0.164\arcm}
\setsymbol{LFI:FWHM:uncertainty:LFI23:units}{0.171\arcm}
\setsymbol{LFI:FWHM:uncertainty:LFI24:units}{0.652\arcm}
\setsymbol{LFI:FWHM:uncertainty:LFI25:units}{1.075\arcm}
\setsymbol{LFI:FWHM:uncertainty:LFI26:units}{1.131\arcm}
\setsymbol{LFI:FWHM:uncertainty:LFI27:units}{1.266\arcm}
\setsymbol{LFI:FWHM:uncertainty:LFI28:units}{1.287\arcm}

\setsymbol{LFI:FWHM:uncertainty:LFI18}{0.170}
\setsymbol{LFI:FWHM:uncertainty:LFI19}{0.174}
\setsymbol{LFI:FWHM:uncertainty:LFI20}{0.170}
\setsymbol{LFI:FWHM:uncertainty:LFI21}{0.156}
\setsymbol{LFI:FWHM:uncertainty:LFI22}{0.164}
\setsymbol{LFI:FWHM:uncertainty:LFI23}{0.171}
\setsymbol{LFI:FWHM:uncertainty:LFI24}{0.652}
\setsymbol{LFI:FWHM:uncertainty:LFI25}{1.075}
\setsymbol{LFI:FWHM:uncertainty:LFI26}{1.131}
\setsymbol{LFI:FWHM:uncertainty:LFI27}{1.266}
\setsymbol{LFI:FWHM:uncertainty:LFI28}{1.287}

\setsymbol{HFI:center:frequency:100GHz:units}{100\,GHz}
\setsymbol{HFI:center:frequency:143GHz:units}{143\,GHz}
\setsymbol{HFI:center:frequency:217GHz:units}{217\,GHz}
\setsymbol{HFI:center:frequency:353GHz:units}{353\,GHz}
\setsymbol{HFI:center:frequency:545GHz:units}{545\,GHz}
\setsymbol{HFI:center:frequency:857GHz:units}{857\,GHz}

\setsymbol{HFI:center:frequency:100GHz}{100}
\setsymbol{HFI:center:frequency:143GHz}{143}
\setsymbol{HFI:center:frequency:217GHz}{217}
\setsymbol{HFI:center:frequency:353GHz}{353}
\setsymbol{HFI:center:frequency:545GHz}{545}
\setsymbol{HFI:center:frequency:857GHz}{857}

\setsymbol{HFI:Ndetectors:100GHz}{8}
\setsymbol{HFI:Ndetectors:143GHz}{11}
\setsymbol{HFI:Ndetectors:217GHz}{12}
\setsymbol{HFI:Ndetectors:353GHz}{12}
\setsymbol{HFI:Ndetectors:545GHz}{3}
\setsymbol{HFI:Ndetectors:857GHz}{4}

\setsymbol{HFI:FWHM:Maps:100GHz:units}{9\parcm88}
\setsymbol{HFI:FWHM:Maps:143GHz:units}{7\parcm18}
\setsymbol{HFI:FWHM:Maps:217GHz:units}{4\parcm87}
\setsymbol{HFI:FWHM:Maps:353GHz:units}{4\parcm65}
\setsymbol{HFI:FWHM:Maps:545GHz:units}{4\parcm72}
\setsymbol{HFI:FWHM:Maps:857GHz:units}{4\parcm39}
\setsymbol{HFI:FWHM:Maps:100GHz}{9.88}
\setsymbol{HFI:FWHM:Maps:143GHz}{7.18}
\setsymbol{HFI:FWHM:Maps:217GHz}{4.87}
\setsymbol{HFI:FWHM:Maps:353GHz}{4.65}
\setsymbol{HFI:FWHM:Maps:545GHz}{4.72}
\setsymbol{HFI:FWHM:Maps:857GHz}{4.39}

\setsymbol{HFI:beam:ellipticity:Maps:100GHz}{1.15}
\setsymbol{HFI:beam:ellipticity:Maps:143GHz}{1.01}
\setsymbol{HFI:beam:ellipticity:Maps:217GHz}{1.06}
\setsymbol{HFI:beam:ellipticity:Maps:353GHz}{1.05}
\setsymbol{HFI:beam:ellipticity:Maps:545GHz}{1.14}
\setsymbol{HFI:beam:ellipticity:Maps:857GHz}{1.19}

\setsymbol{HFI:FWHM:Mars:100GHz:units}{9\parcm37}
\setsymbol{HFI:FWHM:Mars:143GHz:units}{7\parcm04}
\setsymbol{HFI:FWHM:Mars:217GHz:units}{4\parcm68}
\setsymbol{HFI:FWHM:Mars:353GHz:units}{4\parcm43}
\setsymbol{HFI:FWHM:Mars:545GHz:units}{3\parcm80}
\setsymbol{HFI:FWHM:Mars:857GHz:units}{3\parcm67}

\setsymbol{HFI:FWHM:Mars:100GHz}{9.37}
\setsymbol{HFI:FWHM:Mars:143GHz}{7.04}
\setsymbol{HFI:FWHM:Mars:217GHz}{4.68}
\setsymbol{HFI:FWHM:Mars:353GHz}{4.43}
\setsymbol{HFI:FWHM:Mars:545GHz}{3.80}
\setsymbol{HFI:FWHM:Mars:857GHz}{3.67}

\setsymbol{HFI:beam:ellipticity:Mars:100GHz}{1.18}
\setsymbol{HFI:beam:ellipticity:Mars:143GHz}{1.03}
\setsymbol{HFI:beam:ellipticity:Mars:217GHz}{1.14}
\setsymbol{HFI:beam:ellipticity:Mars:353GHz}{1.09}
\setsymbol{HFI:beam:ellipticity:Mars:545GHz}{1.25}
\setsymbol{HFI:beam:ellipticity:Mars:857GHz}{1.03}

\setsymbol{HFI:CMB:relative:calibration:100GHz}{$\lsim 1\%$}
\setsymbol{HFI:CMB:relative:calibration:143GHz}{$\lsim 1\%$}
\setsymbol{HFI:CMB:relative:calibration:217GHz}{$\lsim 1\%$}
\setsymbol{HFI:CMB:relative:calibration:353GHz}{$\lsim 1\%$}
\setsymbol{HFI:CMB:relative:calibration:545GHz}{}
\setsymbol{HFI:CMB:relative:calibration:857GHz}{}

\setsymbol{HFI:CMB:absolute:calibration:100GHz}{$\lsim 2\%$}
\setsymbol{HFI:CMB:absolute:calibration:143GHz}{$\lsim 2\%$}
\setsymbol{HFI:CMB:absolute:calibration:217GHz}{$\lsim 2\%$}
\setsymbol{HFI:CMB:absolute:calibration:353GHz}{$\lsim 2\%$}
\setsymbol{HFI:CMB:absolute:calibration:545GHz}{}
\setsymbol{HFI:CMB:absolute:calibration:857GHz}{}

\setsymbol{HFI:FIRAS:gain:calibration:accuracy:statistical:100GHz}{}
\setsymbol{HFI:FIRAS:gain:calibration:accuracy:statistical:143GHz}{}
\setsymbol{HFI:FIRAS:gain:calibration:accuracy:statistical:217GHz}{}
\setsymbol{HFI:FIRAS:gain:calibration:accuracy:statistical:353GHz}{2.5\%}
\setsymbol{HFI:FIRAS:gain:calibration:accuracy:statistical:545GHz}{1\%}
\setsymbol{HFI:FIRAS:gain:calibration:accuracy:statistical:857GHz}{0.5\%}

\setsymbol{HFI:FIRAS:gain:calibration:accuracy:systematic:100GHz}{}
\setsymbol{HFI:FIRAS:gain:calibration:accuracy:systematic:143GHz}{}
\setsymbol{HFI:FIRAS:gain:calibration:accuracy:systematic:217GHz}{}
\setsymbol{HFI:FIRAS:gain:calibration:accuracy:systematic:353GHz}{}
\setsymbol{HFI:FIRAS:gain:calibration:accuracy:systematic:545GHz}{7\%}
\setsymbol{HFI:FIRAS:gain:calibration:accuracy:systematic:857GHz}{7\%}

\setsymbol{HFI:FIRAS:zero:point:accuracy:100GHz:units}{0.8\MJysr}
\setsymbol{HFI:FIRAS:zero:point:accuracy:143GHz:units}{}
\setsymbol{HFI:FIRAS:zero:point:accuracy:217GHz:units}{}
\setsymbol{HFI:FIRAS:zero:point:accuracy:353GHz:units}{1.4\MJysr}
\setsymbol{HFI:FIRAS:zero:point:accuracy:545GHz:units}{2.2\MJysr}
\setsymbol{HFI:FIRAS:zero:point:accuracy:857GHz:units}{1.7\MJysr}

\setsymbol{HFI:FIRAS:zero:point:accuracy:100GHz}{0.8}
\setsymbol{HFI:FIRAS:zero:point:accuracy:143GHz}{}
\setsymbol{HFI:FIRAS:zero:point:accuracy:217GHz}{}
\setsymbol{HFI:FIRAS:zero:point:accuracy:353GHz}{1.4}
\setsymbol{HFI:FIRAS:zero:point:accuracy:545GHz}{2.2}
\setsymbol{HFI:FIRAS:zero:point:accuracy:857GHz}{1.7}

\setsymbol{HFI:unit:conversion:100GHz:units}{0.2415\MJysrmK}
\setsymbol{HFI:unit:conversion:143GHz:units}{0.3694\MJysrmK}
\setsymbol{HFI:unit:conversion:217GHz:units}{0.4811\MJysrmK}
\setsymbol{HFI:unit:conversion:353GHz:units}{0.2883\MJysrmK}
\setsymbol{HFI:unit:conversion:545GHz:units}{0.05826\MJysrmK}
\setsymbol{HFI:unit:conversion:857GHz:units}{0.002238\MJysrmK}

\setsymbol{HFI:unit:conversion:100GHz}{0.2415}
\setsymbol{HFI:unit:conversion:143GHz}{0.3694}
\setsymbol{HFI:unit:conversion:217GHz}{0.4811}
\setsymbol{HFI:unit:conversion:353GHz}{0.2883}
\setsymbol{HFI:unit:conversion:545GHz}{0.05826}
\setsymbol{HFI:unit:conversion:857GHz}{0.002238}

\setsymbol{HFI:colour:correction:alpha=-2:V1.01:100GHz}{0.9893}
\setsymbol{HFI:colour:correction:alpha=-2:V1.01:143GHz}{0.9759}
\setsymbol{HFI:colour:correction:alpha=-2:V1.01:217GHz}{1.0007}
\setsymbol{HFI:colour:correction:alpha=-2:V1.01:353GHz}{1.0028}
\setsymbol{HFI:colour:correction:alpha=-2:V1.01:545GHz}{1.0019}
\setsymbol{HFI:colour:correction:alpha=-2:V1.01:857GHz}{0.9889}

\setsymbol{HFI:colour:correction:alpha=0:V1.01:100GHz}{1.0008}
\setsymbol{HFI:colour:correction:alpha=0:V1.01:143GHz}{1.0148}
\setsymbol{HFI:colour:correction:alpha=0:V1.01:217GHz}{0.9909}
\setsymbol{HFI:colour:correction:alpha=0:V1.01:353GHz}{0.9888}
\setsymbol{HFI:colour:correction:alpha=0:V1.01:545GHz}{0.9878}
\setsymbol{HFI:colour:correction:alpha=0:V1.01:857GHz}{1.0014}

\newcommand{\msun}{\rm M_{\odot}}
\newcommand{\wjjj}[6]
{{
\left( 
\begin{array}{lcr} #1 & #2 & #3 \\#4 & #5 & #6 \end{array}
\right) 
}}

\def\ben{\begin{enumerate}}
\def\een{\end{enumerate}}
\def\bi{\begin{itemize}}
\def\ei{\end{itemize}}
\def\be{\begin{equation}}                                                                                                          
\def\ee{\end{equation}}
\def\bea{\begin{eqnarray}}                                                                                                         
\def\eea{\end{eqnarray}}

\begin{document}

\title{Cross-correlation of cosmic far-infrared background anisotropies with large scale structures\thanks{Based on observations obtained with \Planck\ (\url{http://www.esa.int/Planck}), an ESA science mission with instruments and contributions directly funded by ESA Member States, NASA, and Canada.}}

\author{P. Serra \inst{1} \and G. Lagache \inst{1} \and O. Dor\'e \inst{2,3} \and A. Pullen \inst{2,3} \and M. White\inst{4,5}} 

\institute{Institut d'Astrophysique Spatiale (IAS), B\^atiment 121, F- 91405 Orsay (France); Universit\'e Paris-Sud 11 and CNRS (UMR 8617) e-mail : pserra@ias.u-psud.fr \and Jet Propulsion Laboratory, California Institute of Technology, 4800 Oak Grove Drive, Pasadena, California, U.S.A. \and California Institute of Technology, Pasadena, California, U.S.A. \and Department of Physics, University of California Berkeley, CA 94720, USA \and Lawrence Berkeley National Laboratory, 1 Cyclotron Rd, Berkeley, CA 94720, USA
\\
}

\date{Received ? / Accepted ?}
\titlerunning{Cross-correlation of Cosmic Far-Infrared Background anisotropies with Large Scale Structures}

\authorrunning{P. Serra et al}
\abstract{We measure the cross-power spectra between luminous red galaxies (LRGs) from the Sloan Digital Sky Survey (SDSS)-III Data Release Eight (DR8) and Cosmic Infrared Background (CIB) anisotropies from \Planck\ and data from the Improved Reprocessing (IRIS) of the Infrared Astronomical Satellite (IRAS) at $353$, $545$, $857$, and $3000$\,GHz, corresponding to $850$, $550$, $350$ and $100\,\mu$m, respectively, in the multipole range $\mathrm{100<\mathrm{l}<1000}$. Using approximately $6.5\cdot10^5$ photometrically determined LRGs in $7760$ deg$^2$ of the northern hemisphere in the redshift range $\mathrm{0.45 < z < 0.65}$, we model the far-infrared background (FIRB) anisotropies with an extended version of the halo model. With these methods, we confirm the basic picture obtained from recent analyses of FIRB anisotropies with {\it Herschel} and \Planck\, that the most efficient halo mass at hosting star forming galaxies is $\mathrm{log(M_{\rm eff}/M_{\odot})=12.84\pm0.15}$. We estimate the percentage of FIRB anisotropies correlated with LRGs as approximately $11.8\%$, $3.9\%$, $1.8\%$, and $1.0\%$ of the total at $3000$, $857$, $545$, and $353$ GHz, respectively. At redshift $\mathrm{z\sim0.55}$, the bias of FIRB galaxies with respect to the dark matter density field has the value $\mathrm{b_{FIRB}\sim1.45}$, and the mean dust temperature of FIRB galaxies is $\rm T_d$=26\,K. Finally, we discuss the impact of present and upcoming cross-correlations with far-infrared background anisotropies on the determination of the global star formation history and the link between galaxies and dark matter.}

\maketitle

\section{Introduction}
The cosmic infrared background (CIB), detected with both the Far Infrared Absolute Spectrophotometer (FIRAS, \citealt{Puget1996,Fixsen1998,Lagache1999}) and the Diffuse Infrared Background Experiment (DIRBE, \citealt{Hauser1998,Lagache2000}), accounts for approximately half of the total energy radiated by structure formation processes throughout cosmic history since the end of the matter-radiation decoupling epoch \citep{Dole2006}. Early attempts at measuring CIB fluctuations at near-infrared wavelengths have been performed from degree to sub-arcminute scales using data from COBE/DIRBE \citep{Kash2000}, IRTS/NIRS  \citep{Matsumoto2005}, 2MASS \citep{Kash2002}, and Spitzer-IRAC \citep{Kash2005}. On the other end, fluctuations in the cosmic far-infrared background (FIRB), composed of thermal emission from warm dust enshrouding star-forming regions in galaxies, have been reported using the ISOPHOT instrument aboard the Infrared Space Observatory \citep{Matsuhara2000,Lagache2000b}, the  Spitzer Space Telescope~\citep{Lagache2007}, the BLAST balloon experiment~\citep{Viero2009}, and the Herschel Space Observatory~\citep{Amblard2011}. In the same period, different cosmic microwave background (CMB) experiments extended these detections to longer wavelengths~\citep{Hall2010,Dunkley2011,Reichardt2012}. The \Planck\ early results paper \citealt{planck2011-6.6} measured angular power spectra of CIB anisotropies from arc-minute to degree scales at $217$, $353$, $545$, and $857$ GHz and the recent paper~\citealt{planck2013-pip56} represents its extension and improvement in terms of analysis and interpretation, establishing \Planck\ as a powerful probe of the FIRB clustering. \\
The far-infrared component of the CIB radiation (or far-infrared background, FIRB) is primarily due to dusty, star-forming galaxies (DSFGs). The dust absorbs the optical and ultraviolet stellar radiation and re-emits in the infrared and submillimetre (submm) wavelengths. The rest frame spectral energy distribution (SED) of DSFGs peaks near $100\mu$m, and it moves into the FIR/submm regime as objects at increasing redshifts are observed. Thus, a complete understanding of the star formation history (SFH) in the Universe must involve accurate observations in the FIR/submm wavelength range.\\ 
FIRB anisotropies are a powerful probe of the global SFH. However, because the signal is integrated over all redshifts, it prevents  a detailed investigation of the temporal evolution of DSFGs over cosmic time. Cross-correlation studies are a powerful expedient to remedy this fact. Because DSFGs trace the underlying dark matter field, a certain degree of correlation between the CIB and any other tracer of the dark matter distribution is expected, provided that some overlap in redshifts exists between the tracers. A useful property of such cross-correlation studies is that the measurement can be used to isolate and analyze a small redshift range in one signal (e.g., the CIB) if the other population is limited in redshift (e.g., the LRGs). In addition, the measurement is not prone to systematics that are not correlated between the two datasets, giving thus a strong signal even if each dataset is contaminated by other physical effects. As shown in~\cite{planck2013-pip56} for example, foreground Galactic dust severely limits CIB measurements at the high frequency channels ($\nu\ge 545$ GHz) while CMB anisotropies contaminate the CIB signal at frequencies $\nu\le353$\,GHz; possible approaches to dealing with these foregrounds include their inclusion in the likelihood analysis (e.g., as a power law) or their removal using a tracer of dust and possibly selecting very clean regions of the sky. In any event, the presence of foreground and background contamination greatly complicates the analysis of CIB data. This limitation disappears when cross-correlating CIB maps with catalogs of dark matter tracers not directly correlated with Galactic dust or CMB (e.g., \citealt{planck2013-p13}); in this case, the presence of uncorrelated contaminants only appears in the computation of the uncertainties associated with the measurement. \\
In this paper, we perform a measurement of the cross-correlation between FIRB maps from \Planck\ and maps from the Improved Reprocessing (IRIS) of the Infrared Astronomical Satellite (IRAS) with a galaxy map of luminous red galaxies (LRGs) from SDSS-III Data Release 8 (DR8).
By fixing both the LRGs redshift distribution and their bias with respect to the dark matter field, we will be able to constrain the most efficient dark matter halo mass at hosting star formation in DSFGs, with their SED. The existence of such a characteristic halo mass has been predicted both analytically and with numerical simulations, and it constitutes a critical component that triggers the growth and assembly of stars in galaxies. After comparing our results with recent analyses in the literature, we will outline the role of upcoming cross-correlation studies with many tracers of the dark matter field in multiple redshift bins, in constraining the redshift evolution of the link between dark matter and star formation, thus bringing new insight into the cosmic SFH. A measurement of the cross-correlation between FIRB sources and other tracers of the dark matter field at high redshift can be extremely important in constraining the early star formation history of the Universe and the clustering properties of high-redshift objects. In this regard, it is important to keep in mind that a good knowledge of the redshift distribution of the sources to be cross-correlated with the FIRB is mandatory in order to constrain both the FIRB emissivity and the bias of both tracers. The quasars catalog from the Wide-field Infrared Survey Explorer (WISE, \citealt{wright2010}), whose redshift distribution can be inferred using the method developed in, e.g., \cite{menard2013}, and the spectroscopic quasars from the Baryon Oscillation Spectroscopic Survey (BOSS,~\citealt{Paris2013}) will certainly be important for such studies. However, our very thorough attempt at computing their cross-power spectrum with the FIRB maps has shown the existence of possible systematics that have not been well understood. In particular, when cross-correlating with the spectroscopic BOSS quasar catalog, we find a  strong anti-correlation at large angular scales whose origin, among many possibilities, has not been clearly isolated. On the other hand, the difficulty in selecting objects in the WISE dataset (beyond the approximate method based on color cuts explained in \citealt{wright2010}), does not allow us to clearly interpret the results of the cross-correlation in the context of the halo model. We thus decided not to include these datasets in the present analysis and to defer this kind of study to a future publication.\\
Throughout this paper, we adopt the standard flat $\Lambda$CDM cosmological model
as our fiducial background cosmology, with parameter values derived from the
best-fit model of the CMB power spectrum measured by \Planck\
\citep{planck2013-p11}:
$\{\Omega_{\rm m},\Omega_{\Lambda},\Omega_{\rm b} h^2,\sigma_8,h,n_{\rm s}\}=   
\{0.3175, 0.6825,  0.022068, 0.8344, 0.6711, 0.9624\}$. We also define halos as matter overdense regions with a mean density equal to $200$ times the mean density of the Universe, and we assume a Navarro-Frenck-White (NFW) profile~\citep{1997ApJ...490..493N} with a concentration parameter as in \citealt{cooray2002}. The fitting function of~\citealt{tinker2008} is used for the halo-mass function while sub-halo mass function and halo bias are taken as in~\citealt{tinker2010}.

\section{Data}
\subsection{CMASS catalog} 
\label{CMASS}
\begin{figure}
  \centering
       \includegraphics[width=77mm,clip=true,trim=0.3cm 0cm 0.2cm 0cm]{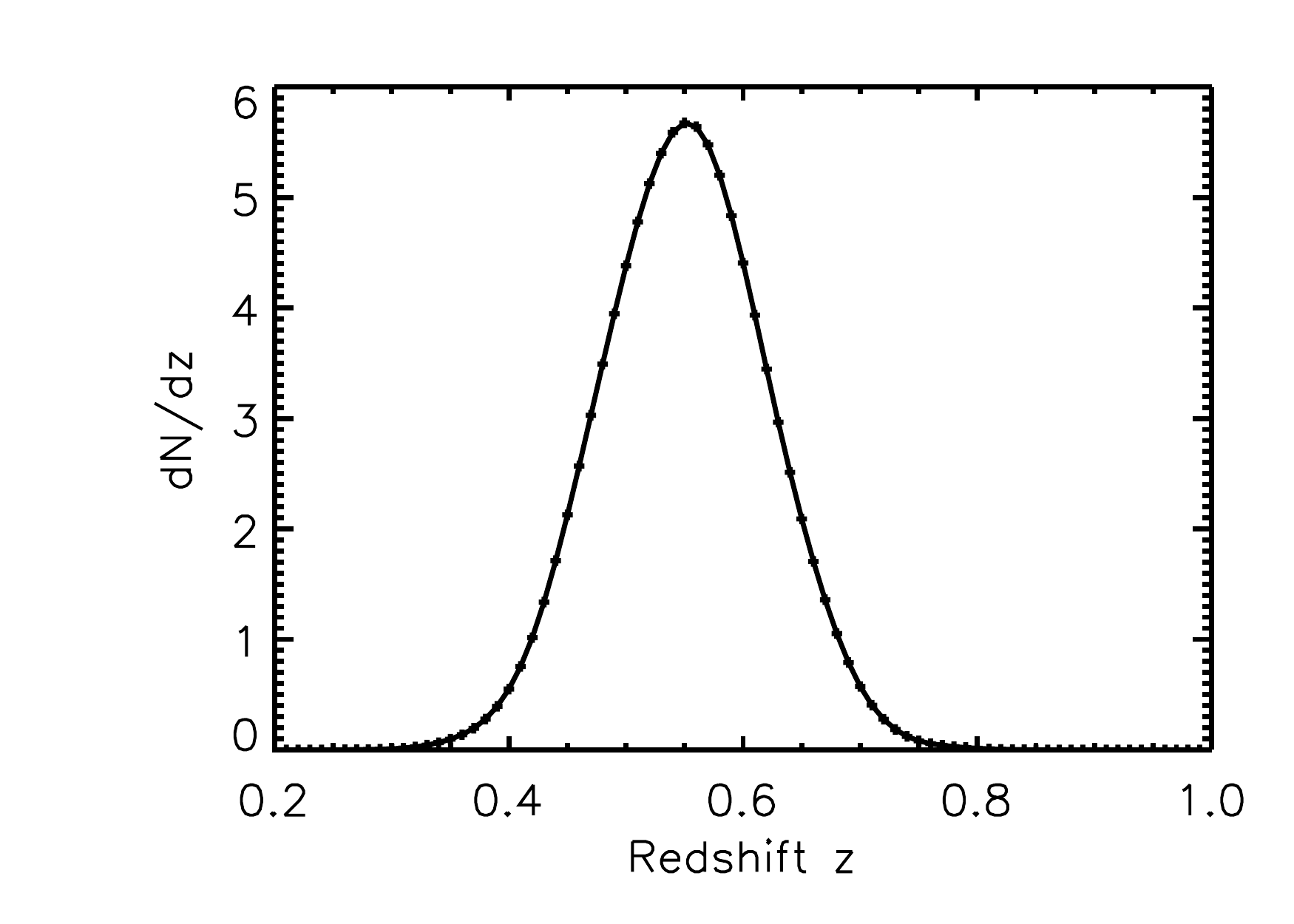}
    \caption[]{Normalized SDSS-III DR8 LRG redshift distribution, peaked at $\bar{z} = 0.55$.}
    \label{LRGdNdz}
 \end{figure}

\begin{figure}
\centering
       \includegraphics[width=90mm,clip=true,trim=0.15cm 0cm 0.2cm 0cm]{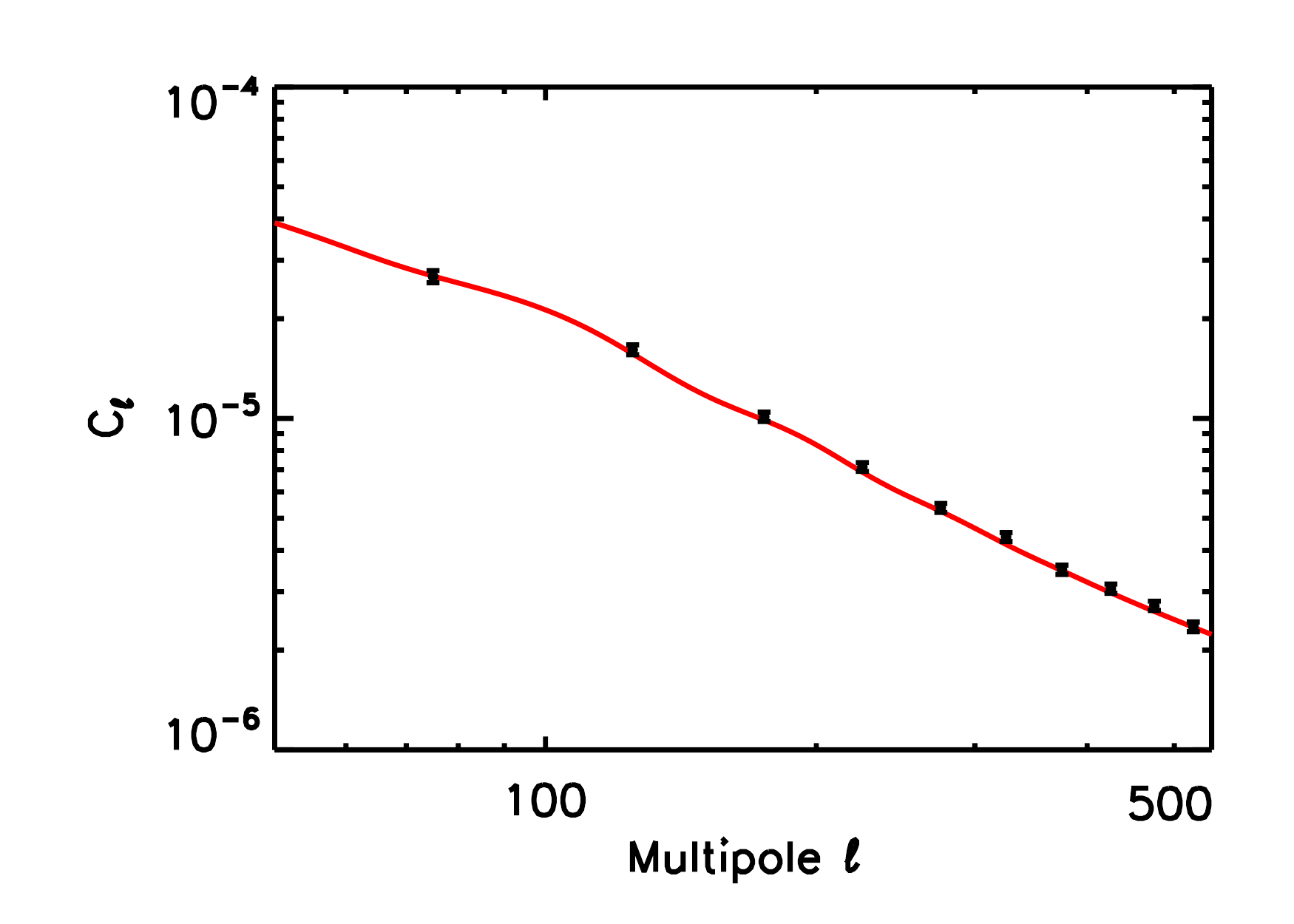}
    \caption[]{Angular auto-power spectrum measured in the SDSS-III DR8 LRG survey}
    \label{LRGCl}
 \end{figure}

The galaxy sample used in the cross-correlation analysis consists of LRGs selected from the publicly available SDSS-III DR8 catalog \footnote{\url{http://portal.nersc.gov/project/boss/galaxy/photoz/}} \citep{Eisenstein2011,Ross2011,Ho2012,dePutter2012}, with photometric redshift in the range $0.45< z_{phot} <0.65$, and centered around $\bar{z}=0.55$ (see Fig~\ref{LRGdNdz}). We considered the same color and magnitude cuts as the SDSS-III ``CMASS'' (constant mass) sample from BOSS~\citep{White2011,Ho2012} and, to create a galaxy map from the catalog, we used the HEALPix\footnote{\url{http://healpix.jpl.nasa.gov/}} pixelization scheme of the sphere \citep{Gorski2005} at resolution Nside $\mathrm{=1024}$; objects are weighted with their probability of being a galaxy and pixelized as number overdensities $\delta_i$ with respect to the mean number of galaxies $\bar{n}$ in each pixel, as:
\begin{equation} 
\delta_i \equiv (n_i-\bar{n)}/\bar{n}.
\end{equation}
Complex survey geometries due to partial sky coverage and masked regions can cause numerical issues and power leakage from large to small scales when performing power spectra computations, especially when using estimators in harmonic space. Because the southern hemisphere footprint has a complicated geometry and it contributes few additional galaxies, to be as conservative as possible, we discard it and only use data in the northern hemisphere. This choice reduces the total area available of approximately $1300$ deg$^2$, leaving $7760$ deg$^2$ with approximately 650,000 galaxies, but ensures stability of results, as we will show below, where we confront error bars computed analytically with those estimated from Monte Carlo simulations. An accurate analysis of potential systematics affecting our dataset, stressing the contribution of seeing effects, sky brightness, and stellar contamination, has been performed in \cite{Ross2011} and \cite{Ho2012}, and we will shape our galaxy mask according to their prescriptions to reduce these effects.\\
As shown in Fig.~\ref{LRGCl}, the computation of the LRG auto-power spectrum from these data is compatible with a non linear prescription for the dark matter power spectrum (we used the Halofit routine \citep{Smith2003} in CAMB\footnote{\url{http://camb.info/}}), together with a scale and redshift independent galaxy bias parameter $b_{\mathrm{LRG}}=2.1\pm0.02$ in agreement with \cite{Ross2011}, and with a galaxy redshift distribution centered in $\bar{z}=0.55$ and with spread $\sigma=0.07$:
\begin{equation}
\frac{dN}{dz}|_{LRG}\propto \mathrm{exp(-(z - \bar{z})^2/(2\sigma^2))},
\label{dNdz}
\end{equation}
as shown in Fig.~\ref{LRGdNdz}. 

In the rest of our analysis, we use Eq.~\ref{dNdz} to compute the LRG redshift distribution and we fix the LRG bias to the value $b_{\mathrm{LRG}}=2.1$.\\

\subsection{\Planck\ and IRIS maps}
\begin{figure}
  \centering
        \includegraphics[width=50mm,angle=90,clip=true,trim=0.05cm 0.cm 0.cm 0.1cm]{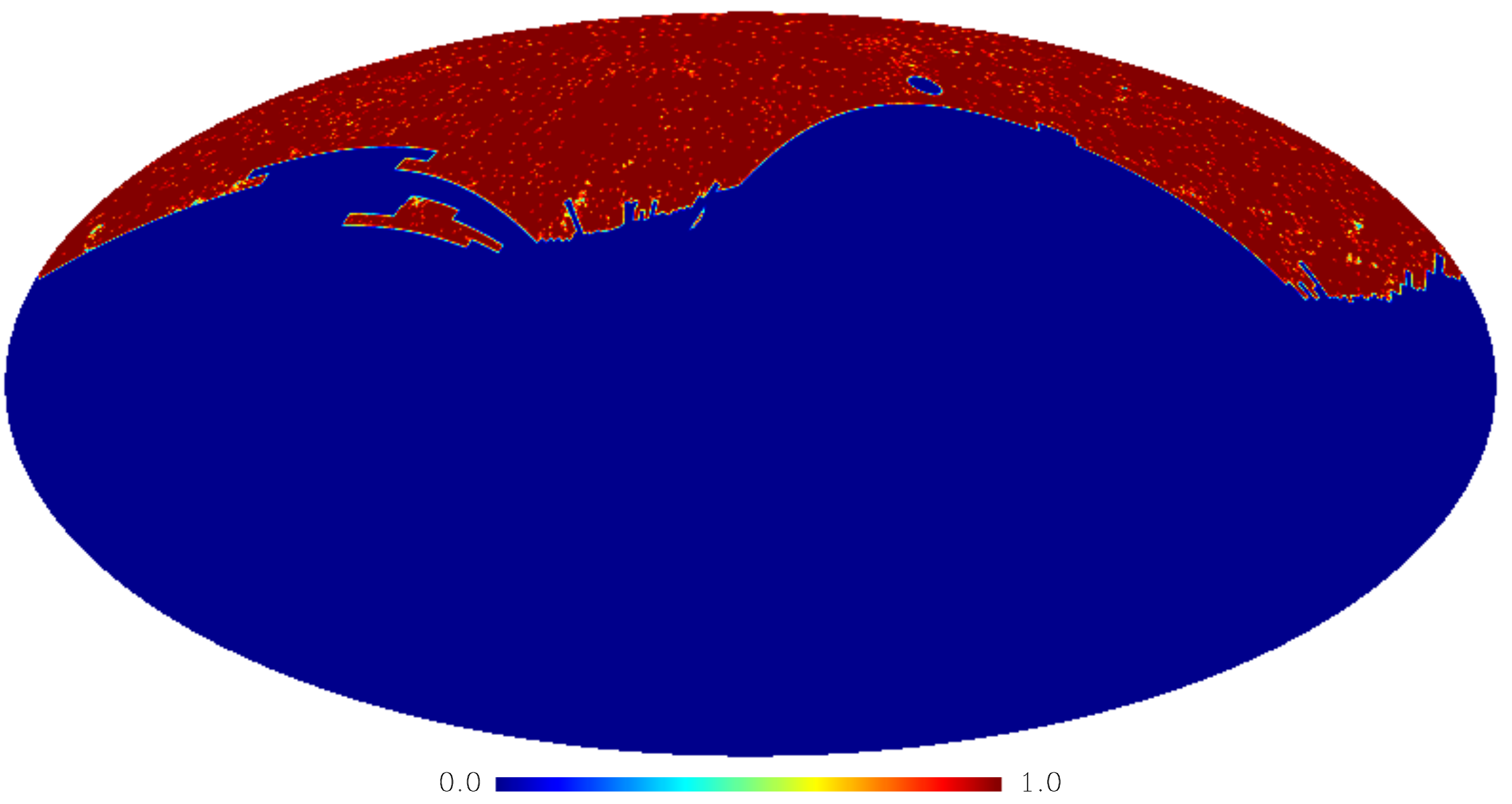}
    \caption{Apodized mask used to cross-correlate the DR8 LRG map with the \Planck\ temperature map at 857 GHz; the sky fraction unmasked is approximately $f_{sky}=0.185$}
    \label{totalmask}
\end{figure}

We use intensity maps at $353$, $545$, and $857$ GHz from the public data release of the first 15.5 months of \Planck\ operations \citep{planck2013-p01}, with a far-infrared map at $3000$ GHz from IRAS (IRIS, \citealt{mivlag2002,mamd2005}). We do not use the lower frequency \Planck\ channels in our analysis, as they contain a large contribution from primary CMB anisotropies and, for the redshift distribution of the galaxy sample considered here, most of the cross-correlation signal comes from the higher frequency maps. 
Finally, the large number of point sources to be masked in the IRAS $3000$ GHz map creates mask irregularities that prevent a stable computation of the cross-power spectrum for multipoles $\ell>500$; we thus consider only measurements at multipoles $\ell<500$ at $3000$ GHz.\\
We refer the reader to \citet{planck2013-p03,planck2013-p03c,planck2013-p03f} for details related to the mapmaking pipeline, beam description and, in general, to the data processing for HFI data. We used two masks to exclude regions with diffuse Galactic emission and extragalactic point sources. The first mask accounts for diffuse Galactic emission as observed in the \Planck\ data and leaves approximately $60\%$ of the sky unmasked \footnote{The mask can be found at \url{http://pla.esac.esa.int/pla/aio/planckResults.jsp?} with $\mathrm{CATEGORY = MASK\_{gal}-06}$}. The second mask has been created using the \Planck\ Catalogue of Compact Sources (PCCS, \citealt{planck2013-p05}) to identify point sources with signal-to-noise ratio greater or equal to five in the maps, and masking out a circular area of $3\sigma$ radius around each source (where $\sigma=FWHM/2.35$). The point sources to be removed have flux densities above a given threshold, as explained in \cite{planck2013-pip56}. At $3000$ GHz, we used a more aggressive mask, which leaves $20\%$ of the sky unmasked and covers dust contaminated regions at high latitudes more efficiently.
The final footprint used in our cross-correlation analysis, which is simply the product of the LRG mask with each of the four FIRB masks, has been smoothed with a Gaussian beam with full width at half maximum of ten arcminutes, to reduce possible power leakage; the mask used for the 857 GHz channel is shown in Fig.~\ref{totalmask}.

\section{Cross-correlation measurement and analysis}
We work in harmonic space, using \texttt{anafast} from the HEALPix package to cross-correlate temperature and density maps and applying the pseudo-$C_{\ell}$ technique described in \cite{Hivon2002} to deconvolve both mask and beam effects from the cross-power spectrum.\\
A generic scalar field $\delta(\hat{n})$ defined over the full-sky can be expressed in terms of spherical harmonics $Y_{lm}(\hat{n})$ as:
\begin{equation}
\delta(\hat{n}) = \sum_{\ell, m}\,a_{\ell m}Y_{\ell m}(\hat{n}),
\end{equation}
where $a_{l,m}$ denotes the spherical harmonic coefficients:
\begin{equation}
a_{\ell m} = \int\,Y_{\ell m}^{*}(\hat{n})\delta(\hat{n})d\Omega
\end{equation}
and, for isotropic temperature and galaxy fields, it is possible to write their cross-power spectrum $C_{\ell}^{Tg}$ as:
\begin{equation}
\langle\,a_{\ell m}^{T}a_{\ell' m'}^{g}\rangle \equiv \langle C_l^{Tg}\rangle\delta_{\ell \ell'}^{K}\delta_{mm'}^{K},
\end{equation}
where $\delta^K$ denotes the Kronecker delta function.\\
Because of contamination or partial sky coverage, we often have access only to a given fraction of the sky. For a generic partial sky map, the resulting power spectrum (called pseudo power spectrum $\hat{C_l}$) is different from the full-sky power spectrum $C_l$, but their ensemble averages are related by:
\begin{equation}
\langle\hat{C_{l}}\rangle = \sum_{l^{'}}M_{l,l^{'}}\langle\,C_{l^{'}}\rangle.
\end{equation}
The coupling matrix $M_{l,l'}$, computed with the mixing matrix formalism introduced in \cite{Hivon2002}, encompasses the combined effects of partial sky coverage, beam, and pixel respectively, and it is obtained by:
\begin{equation}
M_{l,l^{'}} = \frac{2l'+1}{4\pi}\sum_{l_3}
	(2l_3+1) {\mathcal{W}}_{l_3} \wjjj{l}{l'}{l_3}{0}{0}{0}^2 B_{l'}^2,
	\label{eq:kernel_final}
\end{equation}
where we introduced the Wigner 3-$j$ symbol (or Clebsch-Gordan coefficient)
$\wjjj{\ell_1}{\ell_2}{\ell_3}{m_1}{m_2}{m_3}$, and $B_l$ is a window function describing the combined smoothing effect due to the beam and finite pixel size.
\subsection{Error bars computation}
Our measurements are obtained as binned power spectra with a binning $\Delta_\ell=100$, and we use a Monte Carlo approach to compute the uncertainty associated with each bin. In particular, we simulate $\mathrm{N=500}$ pairs of FIRB temperature and LRG density maps, correlated as expected theoretically, adding the expected Poisson noise to both maps, in addition to an instrumental noise and a Galactic dust ``noise'" term to the FIRB frequency maps. More specifically, our pipeline for the computation of error bars, also described in, e.g., \cite{Giannantonio2008}, works as follows:
\begin{itemize}
\item A simulated FIRB frequency map is created (using the program \texttt{synfast} from the HEALPix package) as the sum of a clustering term plus three noise contributions due to shot noise, Galactic dust contamination, and instrument noise. The total power spectrum to be used as input in \texttt{synfast} can be written as follows:
\begin{equation}
C_\ell^{\mathrm{}}(\nu)=C_\ell^{\mathrm{TT,clust}}(\nu)+C_\ell^{\mathrm{TT,SN}}+C_\ell^{\mathrm{TT,dust}}(\nu)+N_\ell^{\mathrm{instr}}(\nu).
\label{eq:corr}
\end{equation}
Using the Limber approximation \citep{Limber1953}, valid on all scales considered in our analysis, the clustering term for each frequency $\nu$ is simply computed as: 
\begin{multline}
C_\ell^{\mathrm{TT,clust}}(\nu) = \int\frac{dz}{\chi^2}\left(\frac{d\chi}{dz}\right)^{-1}b_{FIRB}^2(k,z)\\
\Big(\frac{dS}{dz}(z,\nu)\Big)^2P_{dm}(k=\ell/\chi,z);
\label{eq:auto}
\end{multline}
here $\chi (z)$ is the comoving angular diameter distance to redshift $z$, 
$P_{dm}(k,z)$ is the dark matter power spectrum, 
while $b_{FIRB}$(k,z) and $\frac{dS}{dz}(z,\nu)$ denote the bias of the FIRB sources and the redshift distribution of their emissivity, respectively. Values for the shot-noise power spectrum $C_l^{\mathrm{TT,SN}}$ are obtained from Table~9 of \cite{planck2013-pip56}. For the dust power spectrum we use a template taken as a power law ${C_l^{\mathrm{TT,dust}}\propto\,Kl^{\alpha}}$, where the amplitude $K$ and the slope $\alpha$ are computed by fitting the measured dust-power spectrum from our CIB maps.\\
Finally, the \Planck\ instrument noise power spectrum $N_\ell^{\mathrm{instr}}(\nu)$ is estimated from the jack-knife difference maps, using the first and second halves of each pointing period (see also \citealt{planck2013-pip56}). We refer to \cite{mivlag2002} for the IRAS noise power spectrum computation.
\item A galaxy map is also created as a sum of a clustering plus a shot-noise term:
\begin{equation}
C_\ell^{gg}=C_\ell^{gg,\mathrm{clust}}+C_\ell^{gg,\mathrm{SN}}.
\end{equation}  
In the Limber approximation, the clustering term can be expressed as:
\begin{equation}
C_l^{gg,\mathrm{clust}}=\int\,dz\frac{b_{LRG}^2}{\chi^2(z)} \left( \frac{dN}{dz}(z) \right)^2 P_{dm}(k=l/\chi(z),z)\,,
\end{equation}
and the shot noise power spectrum $C_l^{gg,\mathrm{SN}}$ is directly estimated from the measured 
number of galaxies per pixel.\\
The galaxy map must be correlated with the FIRB map. In general, two correlated galaxy-temperature maps are described by three power spectra, $C_l^{\mathrm{TT}}$, $C_l^{gg}$, and $C_l^{Tg}$, where the last term is given by:
\begin{multline}
C_\ell^{gT}(\nu) = \int\frac{dz}{\chi^2}\left(\frac{d\chi}{dz}\right)^{-1}b_{LRG}
b_{FIRB}(k,z)\frac{dN}{dz}(z)\\
\frac{dS}{dz}(z,\nu)P_{dm}(k=\ell/\chi,z);
\label{eq:cross}
\end{multline}
It is easy to correlate a galaxy map with a given FIRB map created with \texttt{synfast}. First, we build a FIRB map with a power spectrum $C_\ell^{\mathrm{TT}}(\nu)$; then, we make a second map with the same \texttt{synfast} seed used for the clustering term $C_\ell^{\mathrm{TT,clust}}$ and with power spectrum 
$(C_\ell^{\mathrm{Tg}})^2/C_\ell^{\mathrm{TT}}$ and we add this second map to a third map made with a new seed and with power spectrum $C_l^{\mathrm{TT}}-(C_\ell^{\mathrm{Tg}})^2/C_\ell^{\mathrm{TT}}$. These two maps will have amplitudes:
\begin{eqnarray}
a_{\ell m}^{\mathrm{TT}}  &=& \xi_a (C_\ell^{\mathrm{TT}})^{1/2}\\
\nonumber
a_{\ell m}^{gg}  &=& \xi_{a}C_\ell^{\mathrm{Tg}}/(C_\ell^{\mathrm{TT}})^{1/2}+\xi_{b}(C_\ell^{gg}-(C_\ell^{\mathrm{Tg}})^2/C_\ell^{\mathrm{TT}})^{1/2},
\end{eqnarray}
where $\xi$ denotes a random amplitude, which is a complex number with zero mean and unit variance ($\langle\xi\xi^*\rangle=1$ and $\langle\xi\rangle=0$). These amplitudes yield:
\begin{eqnarray}
\langle\,a_{lm}^{\mathrm{TT}}a_{lm}^{\mathrm{TT*}}\rangle&=&C_l^{\mathrm{TT}},\\
\nonumber
\langle\,a_{lm}^{\mathrm{TT}}a_{lm}^{gg*}\rangle&=&C_l^{\mathrm{Tg}},\\
\nonumber
\langle\,a_{lm}^{gg}a_{lm}^{gg*}\rangle&=&C_l^{gg}.
\end{eqnarray}
The obtained maps are then masked with the same mask as that we used to analyze the real data and the cross-power spectrum is then computed using the pseudo-power spectrum technique as in \cite{Hivon2002}.
The set of realizations of the cross-power spectrum provides the uncertainty in our estimate. The covariance matrix of the binned power spectrum $C_b$ is:
\begin{equation}
C_{b,b'}=\langle (C_b-\langle\,C_b\rangle_{\mathrm{MC}})(C_{b'}-\langle\,C_{b'}\rangle_{\mathrm{MC}}\rangle_{\mathrm{MC}} 
\end{equation}
with $\langle\,\cdot\rangle$ standing for Monte Carlo averaging. The error bars on each binned $C_b$ is:
\begin{equation}
\sigma_{C_{b}}=(C_{bb})^{1/2}. 
\end{equation}
The error bars computed from simulations have been also compared with an analytic estimate of the uncertainty, given by:
\begin{equation}
\label{eq:ana_error}
\sigma_{C_{b}}=\left( \frac{1}{(2\ell+1)\Delta\,l} \right)^{1/2} \left[  (C_b^{Tg})^2+(C_b^{gg})(C_b^{\mathrm{TT}})   \right]^{1/2}
\end{equation}
where the term $\mathrm{C_b^{TT}}$ includes power spectra of the FIRB anisotropies, Galactic dust, shot noise, and instrument noise, as: 
\begin{equation}
C_b^{\mathrm{TT}}=C_b^{\mathrm{TT,FIRB}}+C_b^{\mathrm{dust}}+C_b^{\mathrm{SN}} + N_b^{\mathrm{instr}}.
\label{eq:analytical}
\end{equation}
For each frequency considered, our Monte Carlo estimates of the uncertainties are within $10\%$ of the uncertainties derived from Eq.~\ref{eq:ana_error}.  
In the fitting process, we thus conservatively increase our simulated error bars by $10\%$. We have also checked that cross-correlating simulated maps created from different input power spectra and masked in different ways, we are always able to retrieve the input spectra, within statistical uncertainties, ensuring the stability of our results. 
\end{itemize}
\subsection{Null tests}
\begin{figure}
  \includegraphics[width=90mm]{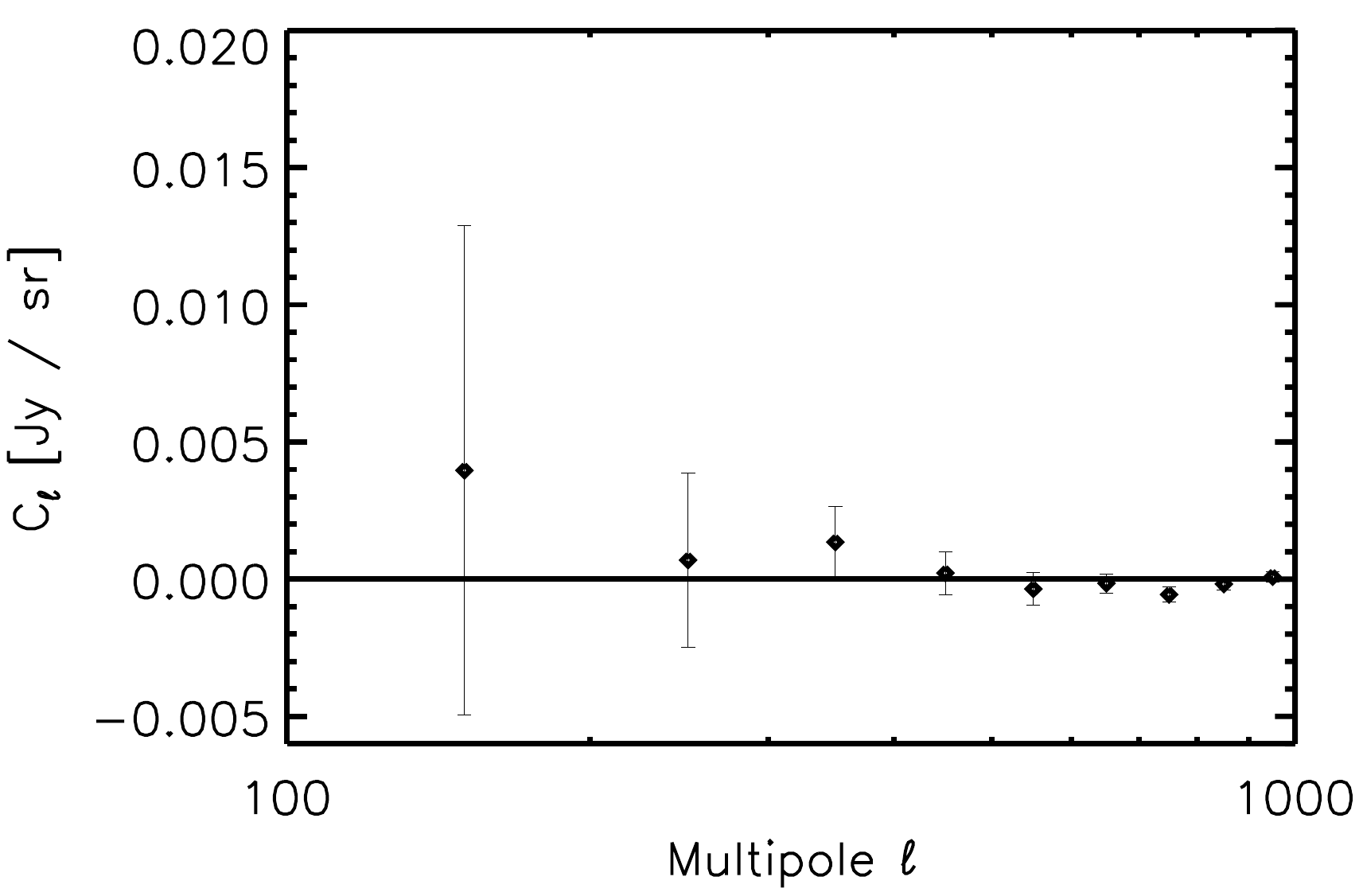}
    \caption{Mean cross-power spectrum of 500 FIRB random maps at $857$\,GHz with the LRG map; the signal is perfectly compatible with zero.}
    \label{Null}
\end{figure}

In order to test for possible contaminants in our datasets, we also performed two null tests. In the first test, we cross-correlated 500 FIRB temperature random maps at $857$\,GHz (adding the expected level of foreground dust and instrumental noise) with the LRG map. The mean of the cross-correlation signal and its uncertainty are plotted in Fig.~\ref{Null}; with a $\chi^2$ of $6.7$ for $9$ degrees of freedom, our p-value is $0.67$ and the null-test hypothesis of correlation consistent with zero is accepted.\\ 
We also performed a rotation test \citep{sawangwit2010,giannantonio2012}, where one of the maps is rotated by an arbitrary angle and then cross-correlated with the other map: if the rotation angle $\Delta\phi$ is large enough, and in absence of systematics, the resulting cross-power spectrum should be compatible with zero. Keeping the FIRB map mixed, we computed $\mathrm{N=89}$ cross-power spectra with N galaxy maps, rotated by $\Delta\phi=4$ degrees with respect to each other and used the corresponding rotated galaxy masks; with $\chi^2=16.5$ for 9 degrees of freedom, we accept the null-test hypothesis of correlation consistent with zero.  
\subsection{Analysis and results}
\begin{figure*}
  \begin{center}
    \includegraphics{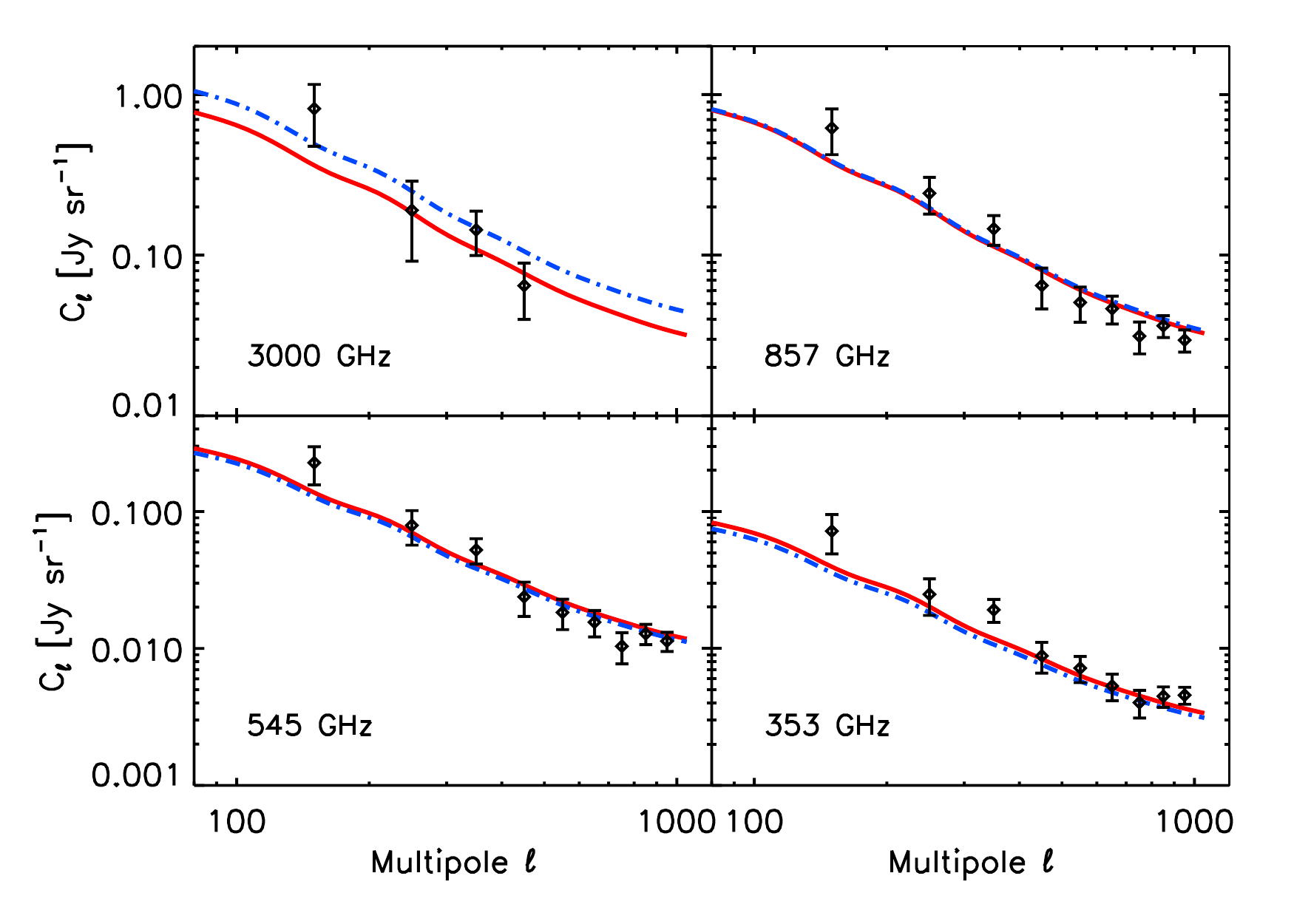}
    \caption{Cross-power spectra measurements among CIB maps at $3000$, $857$, $545$, and $353$ GHz, and CMASS LRGs (black points); for the IRIS $3000$ GHz channel we did not use data points at multipoles $\mathrm{l>500}$. Best-fit power spectra are shown for the parametric SED functional form (red lines) and for the effective SED (blue dashed-dotted lines).}
    \label{power_spectra}
  \end{center}
\end{figure*}

In Fig.~\ref{power_spectra}, we show the cross-power spectra measured for the four frequencies considered and with the uncertainty computed from Monte Carlo simulations. The statistical significance of the signal is obtained by summing, for each frequency $\nu$, the significance in the different multipole bins i as
\begin{equation}
s_{\nu} = \sqrt{\sum_{i=1}\Big(\frac{C_i}{\sigma_{C_i}}\Big)^2};
\end{equation}  
we obtain values of $8.7$, $13.0$, $12.3$, and $12.6$ at $3000$, $857$, $545$, and $353$ GHz, respectively.

The theoretical cross-power spectrum is given by Eq.~\ref{eq:cross} and a Poisson term, because of far-infrared emission from individual galaxies, is not included, because for the LRG's number density and average FIR emission, it is negligible. The redshift distribution of FIRB sources at the observed frequency 
$\nu$ is connected to the mean FIRB emissivity per comoving unit 
volume $j_{\nu}(z)$ through the relation
\begin{equation}
\frac{dS_{\nu}}{dz} = \frac{c}{H(z)(1+z)}\bar{j}_{\nu}(z),
\label{eq:emiss}
\end{equation}
where the galaxy emissivity $\bar{j}_{\nu}(z)$ can be written as           
\begin{equation}
\bar{j}_{\nu}(z) = \int dL \frac{dn}{dL}(L,z)
\frac{L_{(1+z)\nu}}{4\pi};
\label{eq_jnu}
\end{equation}
here $L_{(1+z)\nu}$ and $dn/dL$ denote the infrared galaxy luminosity and luminosity function, respectively, while the term $\mathrm{(1+z)\nu}$ denotes the rest-frame frequency. 
The emissivity $j_\nu(z)$ is modeled with a halo-model approach, introduced in \cite{shang2012} and successfully applied in, e.g., \cite{planck2013-pip56,viero2012}, whose main feature is the introduction of a parametric form to describe the dependence of the galaxy luminosity on its host halo mass. This allows us to overcome the unrealistic assumption, typical of many models based on a set of infrared luminosity functions and a prescription to populate galaxies in dark matter halos using a halo occupation distribution (HOD) formalism that galaxies have all the same luminosity and contribute to the emissivity density in the same way, despite their dark matter environment. \cite{shang2012} provide a detailed description of the theoretical motivations and limitations of the modeling.\\
In this model, the galaxy infrared luminosity $L_{(1+z)\nu}$ is linked to the host dark matter halo mass using the following parametric form:
\begin{equation}
L_{(1+z)\nu}(M,z)=L_0 \left(1+z\right)^{3.6} \Sigma(M) \Theta[(1+z)\nu]
\label{eqn:lfunc}
\end{equation}
where the redshift-dependent, global normalization, has been fixed to the mean value found in \cite{planck2013-pip56}, while the term $L_0$ is a normalization parameter constrained with the CIB mean level at the frequencies considered. We will not discuss this parameter further in the rest of our analysis. \\ 
We also assume a log-normal function $ \Sigma(M)$ for the dependence of the galaxy
luminosity on halo mass
 \begin{equation}
\Sigma (M)=  \frac{M}{(2\pi \sigma_{L/M}^2)^{1/2}}\,
e^{-(\log_{10}(M)-\log_{10}(M_{\rm eff}))^2/                                    
 2\sigma_{L/M}^2}
\label{eqn:sigmam}
\end{equation}
where $M_{\rm eff}$ and $\sigma_{L/M}$ describe the peak of the specific IR emissivity and the range of halo masses which is more efficient at producing star formation; following \cite{shang2012,planck2013-pip56}, we assume the condition $\sigma^2_{L/M}=0.5$ while we fix the minimum halo mass at $\mathrm{M_{min}}=10^{10}$$\mathrm{M_{\odot}}$ (compatible with \cite{viero2012}) throughout this paper. The log-normal functional form used here describes the observation that a limited range of halo masses dominates the star formation activity. Recent cosmological simulations suggest that processes such as photoionization, supernovae heating, feedback from active galactic nuclei, and virial shocks suppress star formation at both the low- and the high-mass end \citep{Benson2003,Croton2006,Silk2003,Bertone2005,Birn2003,Keres2005,Dekel2006}; it is thus possible to introduce a characteristic mass scale $M_{\rm eff}$, which phenomenologically illustrates the impact of dark matter on star formation processes.\\
In general, a modified back body functional form
\citep[see][and reference therein]{blain2002} can be assumed for galaxy SEDs            
\begin{equation}
\Theta (\nu,z) \propto
\left\{\begin{array}{ccc}
\nu^{\beta}B_{\nu}\,(T_{\rm d})&
 \nu<\nu_0\, ;\\
\nu^{-\gamma}&  \nu\ge \nu_0\, 
\end{array}\right.
\label{eqn:thetanu}
\end{equation}
where $B_{\nu}$ denotes the Planck function, while the emissivity index
$\beta$ gives information about the physical
nature of dust, in general depending on grain composition,
temperature distribution of tunneling states and wavelength-dependent
excitation \citep[e.g.,][]{meny2007}. 
The power-law function is used to temper the exponential (Wien) tail
at high frequencies and obtain a shallower SED shape, which is more in agreement
with observed SEDs (see, e.g.,~\citealt{blain2002}). The two SED functions at high and low frequencies are connected smoothly at the
frequency $\nu_0$ satisfying
\begin{equation}
\frac{d{\rm ln}\Theta(\nu,z)}{d\rm{ln}\nu} =-\gamma.
\end{equation}
We explicitly checked that our data do not allow us to strongly constrain the emissivity index $\beta$ and the SED parameter $\gamma$; thus, we fixed their values to the mean values found by \cite{planck2013-pip56}, as $\beta=1.7$ and $\gamma=1.7$. Finally, the parameter $T_d$ describes the average dust temperature of FIRB sources at $z\sim0.55$. Note that, since our measurement is restricted to quite a narrow redshift bin, we do not consider a possible redshift dependence of parameters such as $M_{\rm eff}$ or $T_d$; the only redshift-dependent quantity is the global normalization term $\Phi (z)$. \\
The parameter space is sampled using a Monte Carlo Markov chain analysis with a modified version of the publicly available code {\tt CosmoMC}
\citep{lewis2002}.
We consider variations in the following set of three halo model parameters:
\begin{equation}
\mathscr{P} \equiv \{M_{\rm eff},T_d,L_0 \};
\end{equation}
we assume the following priors on our physical parameters: $\mathrm{\rm \log(M_{\rm eff})\in[11:13]}{\rm M}_{\odot}$ and $\rm T_{d}\in[20:60]$K, and we explicitly checked that our results do not depend on the priors assumed.

\begin{table}[!tbh]
\begingroup
\newdimen\tblskip \tblskip=5pt
\caption{Mean values and marginalized $68\%$ c.l. for halo model
parameters }
\label{HOD_param_values}
\nointerlineskip
\vskip -3mm
\footnotesize
\setbox\tablebox=\vbox{
 \newdimen\digitwidth
 \setbox0=\hbox{\rm 0}
  \digitwidth=\wd0
  \catcode`*=\active
  \def*{\kern\digitwidth}
  \newdimen\signwidth
  \setbox0=\hbox{+}
  \signwidth=\wd0
  \catcode`!=\active
  \def!{\kern\signwidth}
\halign{\tabskip=0pt#\hfil\tabskip=1.0em&
  #\hfil\tabskip=1.0em&
  \hfil#\hfil\tabskip=0pt\cr
\noalign{\doubleline}
\noalign{\vskip -2pt}
Parameter& Definition& Mean value\cr
\noalign{\vskip 3pt\hrule\vskip 3pt}
$\rm \log(M_{\rm eff}) [{\rm M}_{\odot}]$& Halo model most efficient mass&
 $12.84\pm0.15*$\cr
$\rm T_d$ [K]& SED: dust temperature ($\bar{z}$=0.55)& $26.0 \pm1.3$ \cr
\noalign{\vskip 3pt\hrule\vskip 3pt}}}
\endPlancktablewide
\endgroup
\end{table}

\begin{figure}[!t]
\centering
       \includegraphics[width=90mm,clip=true,trim=0.0cm 0cm 0.cm 0cm]{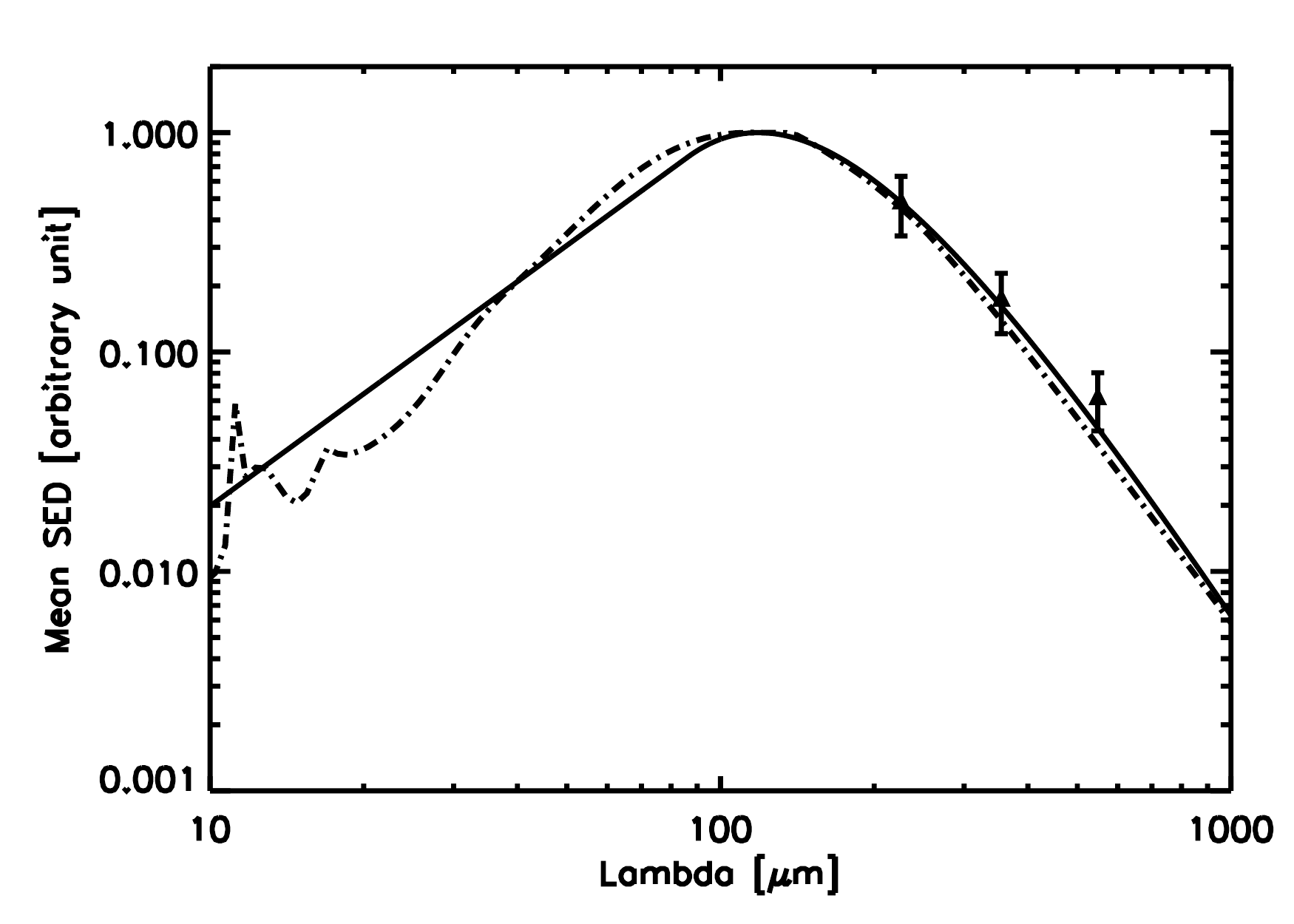}
    \caption{Best-fit SEDs at $z=0.55$ (both normalized to have amplitude equal to one at  $\lambda$=120\,$\rm \mu$m) computed with the modified blackbody functional form (continuous line) and with the mean effective SED approach (dot-dash line), as in \cite{Beth2012}. Data points show a (normalized) estimate for the SED of FIRB fluctuations computed from the ratio of the cross-power spectra at multipole $l=450$.}
    \label{plotSED}
 \end{figure}
In addition to fitting to the four cross-power spectra, we also use the mean level of the CIB at $3000$ GHz ($12.6^{+8.3}_{-1.7}$ nW m$^{-2}$sr$^{-1}$), $857$ GHz ($6.5^{+1.7}_{-1.6}$ nW m$^{-2}$sr$^{-1}$) and $545$ GHz ($2.1^{+0.7}_{-0.6}$ nW m$^{-2}$sr$^{-1}$) deduced from galaxy number counts as useful priors to constrain the global normalization parameter $L_0$.\\
Our set of three parameters allows us to obtain a very good fit to the data (Fig.~\ref{power_spectra}), with a best-fit $\chi^2$ of $26.9$ for $31$ degrees of freedom. Mean values and marginalized limits on the halo-model parameters considered here are listed in Table~\ref{HOD_param_values}. They are in good agreement with results obtained from the analysis of \Planck\ and IRIS auto-power spectra using the same parameterization for the FIRB emissivity in \citealt{planck2013-pip56}, providing a strong confirmation of their results. \\
Using the values in Table~\ref{HOD_param_values}, we also compute the approximate fraction of the total FIRB responsible for the cross-correlation with LRGs, obtaining $11.8\%$ at 3000 GHz, $3.9\%$ at 857 GHz, $1.8\%$ at $545$ GHz, and $1\%$ at 353 GHz. Compatible results are obtained through the computation of the mean coherence \citep{Kash2012b,Cappelluti2012} of the signal, defined as the square of the ratio of the cross-power spectra to the geometric mean of the auto-power spectra:
\begin{equation}
\it{\mathrm{C}}_l^{\nu} = \frac{(C_l^{Tg,\nu})^2}{C_l^{T,\nu}C_l^{G}}.
\end{equation}
Finally, the mean values inferred for both the dust temperature ($\mathrm{T_{\rm d}}=26.0\pm1.3$\,K) and the bias ($b_{FIRB}\sim1.45)$ are compatible with results obtained by \cite{planck2013-pip56} at $z\simeq 0.55$.\\
To test the robustness of our results, we also replaced our parametric SED functional form with the mean effective SED $S_{\nu,eff}$ of all galaxies at redshift $z=0.55$ from \cite{Beth2012}, and we repeated the analysis, keeping only $\mathrm{M_{\rm eff}}$ and $\mathrm{L_0}$ as free parameters. With a $\chi^2_{red}$ of $1.0$ (best-fit $\chi^2$ equal to $31.9$ for $32$ degrees of freedom) our fit is satisfying; however, the most efficient halo mass at hosting star-forming galaxies is $\mathrm{M_{\rm eff}=12.90\pm0.14}$, higher than \cite{planck2013-pip56}, and compatible at $95\%$ c.l.. \\
As we can see in Fig.~\ref{plotSED}, the parametric form used in this paper is similar to the mean effective SED for the range of wavelengths that matters in this analysis, yielding a satisfactory confirmation of the goodness of our results. However, the $\gamma$ parameter, which controls the high frequency tail of the spectrum, fails to correctly reproduce the shape of real galaxy SEDs in the mid-infrared regime (from $10$ to $30$ $\mu$m), mainly because of broad line absorption from polycyclic aromatic hydrocarbon (PAH) molecules. In Fig.~\ref{plotSED}, we also plot the value of the FIRB SED computed from the ratio of the measured cross-power spectra at multipole $\mathrm{l}=450$ and shifted to redshift $z=0.55$; our results are compatible with the best-fit SEDs used in the paper.\\  
On linear scales, the cross-power spectra considered can also be fitted by power laws. Assuming a simple, two-parameter functional form such as
\begin{equation}
C_l = A\left(\frac{l}{100}\right)^n,
\end{equation} 
it is possible to obtain a very good fit at all frequencies over the multipole range $100<l<500$. Mean values and marginalized limits on the amplitudes $A_{\nu}$ and power-law slopes $n_{\nu}$ are provided in Table~\ref{plaw_param_values}.

\begin{table}[!tbh]
\begingroup
\newdimen\tblskip \tblskip=5pt
\caption{Mean values and marginalized $68\%$ c.l. for power-law parameters}
\label{plaw_param_values}
\nointerlineskip
\vskip -3mm
\footnotesize
\setbox\tablebox=\vbox{
 \newdimen\digitwidth
 \setbox0=\hbox{\rm 0}
  \digitwidth=\wd0
  \catcode`*=\active
  \def*{\kern\digitwidth}
  \newdimen\signwidth
  \setbox0=\hbox{+}
  \signwidth=\wd0
  \catcode`!=\active
  \def!{\kern\signwidth}
\halign{\tabskip=0pt#\hfil\tabskip=1.0em&
  #\hfil\tabskip=1.0em&
  \hfil#\hfil\tabskip=0pt\cr
\noalign{\doubleline}
\noalign{\vskip -2pt}
Frequency& A & n\cr
\noalign{\vskip 3pt\hrule\vskip 3pt}
$\rm 353$  & $0.18\pm 0.07 $ & $-1.90 \pm 0.36$\cr
$\rm 545$  & $0.60\pm 0.23$ & $-2.08 \pm 0.37$ \cr
$\rm 857$  & $1.72\pm 0.65$ & $-2.09 \pm 0.36$  \cr
$\rm 3000$ &$2.45\pm 1.33$ & $-2.42\pm 0.57$ \cr
\noalign{\vskip 3pt\hrule\vskip 3pt}}}
\endPlancktablewide
\endgroup
\end{table}

\subsection{Discussion}
As mentioned earlier, the dark matter halo mass influences the evolution of galaxies, driving processes of gas accretion and star formation rate (SFR) (see, e.g., \citealt{Rees1977,Silk1977,White1978,Fall1980,Blu1984}). Numerous techniques have been considered to constrain the link between dark matter halos and their host galaxies (see discussion and references in \citealt{Behroozi2010, Moster2010}). From a theoretical point of view, semi analytic models and hydrodynamical simulations have been developed to study galaxy formation processes ab initio; however, they are still not able to reproduce many observational evidences and suffer from several uncertainties. A different approach, which does not rely on assumptions related to poorly constrained physical processes and is based instead on a statistical description of the link between galaxies and dark matter, is the basis for the HOD formalism and the abundance matching technique, which also assumes the existence of a monotonic relation between halo mass and galaxy stellar mass. Results based on abundance matching, applied to optical and infrared data \citep{Dore2012,behroozi2012,Moster2010,Guo2010}, constrain the characteristic halo mass at $M_{\rm eff}\sim10^{12}\msun$, with little evolution in redshift and some uncertainty, mainly due to systematics in the stellar mass estimates \citep{Behroozi2010}. The higher value obtained in this work ($M_{\rm eff}\sim 6.9\cdot10^{12}\msun$), can be due to the adoption of a log-normal function to model the luminosity-mass relation, while the previously mentioned analyses use a more complicated relation, to take the slow decrease of the stellar mass to halo mass ratio at the high mass end into account. However, the mean value for the most efficient mass found here is in agreement with, or only slightly higher than, results from recent analyses of FIRB anisotropies \citep{shang2012,xia2012,viero2012,planck2013-pip56}. \\
We note that the narrow redshift bin involved in our analysis imposes the use of a very simple interpretative model for both the LRGs and the FIRB galaxies. In particular, both the redshift evolution of the galaxy FIR luminosity and some SED parameters are kept fixed. In an upcoming study, we will cross-correlate FIRB anisotropies with a combination of data from multiple surveys such as WISE, GALEX, and BOSS. This will allow us to probe the link between dark matter and galaxies over a wide redshift range, from low redshifts ($0<z<1$) where a steep evolution of the cosmic star formation has been measured, to high redshifts ($1<z<3$), where the SFR peaks. Introducing redshift-dependent quantities and a realistic scatter in the relation between galaxy luminosity and dark matter halo mass (that has been neglected here, see discussion in \citealt{shang2012}), we will be able to study the temporal evolution of the most efficient halo mass at hosting star formation (possibly constraining downsizing scenarios, \citealt{Cowie1996,Bundy2006,Conroy2009}), thus establishing cross-correlations with FIRB anisotropies as a powerful probe of the link between galaxies and dark matter and the global star formation history. 
\section{Conclusions}
We have measured the cross-correlation between LRGs from SDSS-III DR8 and CIB anisotropies from IRIS and \Planck\ at $3000$, $857$, $545$, and $353$ GHz. Using an extended version of the halo model that connects galaxy luminosity to host halo mass with a simple parametric form, we confirmed the basic results obtained from recent analyses of FIRB anisotropies from {\it Planck} and {\it Herschel}, with a most efficient halo mass at hosting star-forming galaxies constrained at the value $\mathrm{log(M_{\rm eff}/M_{\odot})=12.84\pm0.15}$, and the mean dust temperature of FIR sources at $z\sim0.55$ with the value $\mathrm{T_d=26.0\pm1.3}$K.\\
The cross-correlation of FIRB sources with other tracers of the dark matter field now starts being successfully exploited to constrain the interplay between dark and luminous matter (see, e.g, \citealt{Wang2014}), and cross-correlations with data from present and upcoming surveys are particularly promising for many reasons. First of all, while the measurement of FIRB auto-power spectra will not improve much in the near future, as they are limited by issues related to components separation, cross-correlations with FIRB anisotropies are limited mostly by noise and systematics. In particular, control of the effects of Galactic dust will be critical to achieving a precise measurement of the cross-correlation, since it contributes significantly to the total uncertainty. Dust subtraction, using ancillary data and clean regions of the sky (as done in \citealt{planck2013-pip56}) can improve the signal-to-noise ratio, provided that large enough fields are considered. A careful analysis of systematics is also mandatory to exclude the possibility of correlated systematics between datasets; an example is stellar density affecting quasar catalogs and Galactic dust affecting FIRB anisotropies, which correlate and contaminate the cross-correlation signal.\\
Cross-correlations with FIRB anisotropies are also important because they allow us to isolate the CIB signal in multiple redshift slices and thus constrain the link between galaxies and dark matter and global star formation as a function of redshift. These studies will be particularly interesting when performed with sources at high redshift (such as, e.g., quasars from BOSS), where the cosmic SFR peaks and current model uncertainties for DSFGs are large.

\begin{acknowledgements}
The development of \Planck\ has been supported by: ESA; CNES and CNRS/INSU-IN2P3-INP (France); ASI, CNR, and INAF (Italy); NASA and DoE (USA); STFC and UKSA (UK); CSIC, MICINN and JA (Spain); Tekes, AoF and CSC (Finland); DLR and MPG (Germany); CSA (Canada); DTU Space (Denmark); SER/SSO (Switzerland); RCN (Norway); SFI (Ireland); FCT/MCTES (Portugal); and PRACE (EU). A description of the \Planck\ Collaboration and a list of its members, including the technical or scientific activities in which they have been involved, can be found at \url{http://www.sciops.esa.int/index.php?project=planck&page=Planck_Collaboration}.\\
We would like to thank the anonymous referee for providing us with constructive comments and suggestions. P.S. would like to thank Alex Amblard and Shirley Ho for useful discussions. Part of the research described in this paper was carried out at the Jet Propulsion Laboratory, California Institute of Technology, under a contract with the National Aeronautics and Space Administration.
\end{acknowledgements}

\bibliographystyle{aa}
\bibliography{main_bib_CIB,Planck_bib}

\begin{thebibliography}{78}
\expandafter\ifx\csname natexlab\endcsname\relax\def\natexlab#1{#1}\fi

\bibitem[{{Amblard} {et~al.}(2011){Amblard}, {Cooray}, {Serra}, {Altieri},
  {Arumugam}, {Aussel}, {Blain}, {Bock}, {Boselli}, {Buat},
  {Castro-Rodr{\'{\i}}guez}, {Cava}, {Chanial}, {Chapin}, {Clements}, {Conley},
  {Conversi}, {Dowell}, {Dwek}, {Eales}, {Elbaz}, {Farrah}, {Franceschini},
  {Gear}, {Glenn}, {Griffin}, {Halpern}, {Hatziminaoglou}, {Ibar}, {Isaak},
  {Ivison}, {Khostovan}, {Lagache}, {Levenson}, {Lu}, {Madden}, {Maffei},
  {Mainetti}, {Marchetti}, {Marsden}, {Mitchell-Wynne}, {Nguyen}, {O'Halloran},
  {Oliver}, {Omont}, {Page}, {Panuzzo}, {Papageorgiou}, {Pearson},
  {P{\'e}rez-Fournon}, {Pohlen}, {Rangwala}, {Roseboom}, {Rowan-Robinson},
  {Portal}, {Schulz}, {Scott}, {Seymour}, {Shupe}, {Smith}, {Stevens},
  {Symeonidis}, {Trichas}, {Tugwell}, {Vaccari}, {Valiante}, {Valtchanov},
  {Vieira}, {Vigroux}, {Wang}, {Ward}, {Wright}, {Xu}, \&
  {Zemcov}}]{Amblard2011}
{Amblard}, A., {Cooray}, A., {Serra}, P., {et~al.} 2011, \nat, 470, 510

\bibitem[{{Behroozi} {et~al.}(2010){Behroozi}, {Conroy}, \&
  {Wechsler}}]{Behroozi2010}
{Behroozi}, P.~S., {Conroy}, C., \& {Wechsler}, R.~H. 2010, \apj, 717, 379

\bibitem[{{Behroozi} {et~al.}(2012){Behroozi}, {Wechsler}, \&
  {Conroy}}]{behroozi2012}
{Behroozi}, P.~S., {Wechsler}, R.~H., \& {Conroy}, C. 2012, ArXiv e-prints

\bibitem[{{Benson} {et~al.}(2003){Benson}, {Bower}, {Frenk}, {Lacey}, {Baugh},
  \& {Cole}}]{Benson2003}
{Benson}, A.~J., {Bower}, R.~G., {Frenk}, C.~S., {et~al.} 2003, \apj, 599, 38

\bibitem[{{Bertone} {et~al.}(2005){Bertone}, {Stoehr}, \&
  {White}}]{Bertone2005}
{Bertone}, S., {Stoehr}, F., \& {White}, S.~D.~M. 2005, \mnras, 359, 1201

\bibitem[{{B{\'e}thermin} {et~al.}(2012{\natexlab{a}}){B{\'e}thermin}, {Daddi},
  {Magdis}, {Sargent}, {Hezaveh}, {Elbaz}, {Le Borgne}, {Mullaney}, {Pannella},
  {Buat}, {Charmandaris}, {Lagache}, \& {Scott}}]{Beth2012}
{B{\'e}thermin}, M., {Daddi}, E., {Magdis}, G., {et~al.} 2012{\natexlab{a}},
  \apjl, 757, L23

\bibitem[{{B{\'e}thermin} {et~al.}(2012{\natexlab{b}}){B{\'e}thermin},
  {Dor{\'e}}, \& {Lagache}}]{Dore2012}
{B{\'e}thermin}, M., {Dor{\'e}}, O., \& {Lagache}, G. 2012{\natexlab{b}}, \aap,
  537, L5

\bibitem[{{Birnboim} \& {Dekel}(2003)}]{Birn2003}
{Birnboim}, Y. \& {Dekel}, A. 2003, \mnras, 345, 349

\bibitem[{{Blain} {et~al.}(2003){Blain}, {Barnard}, \& {Chapman}}]{blain2002}
{Blain}, A.~W., {Barnard}, V.~E., \& {Chapman}, S.~C. 2003, \mnras, 338, 733

\bibitem[{{Blumenthal} {et~al.}(1984){Blumenthal}, {Faber}, {Primack}, \&
  {Rees}}]{Blu1984}
{Blumenthal}, G.~R., {Faber}, S.~M., {Primack}, J.~R., \& {Rees}, M.~J. 1984,
  \nat, 311, 517

\bibitem[{{Bundy} {et~al.}(2006){Bundy}, {Ellis}, {Conselice}, {Taylor},
  {Cooper}, {Willmer}, {Weiner}, {Coil}, {Noeske}, \& {Eisenhardt}}]{Bundy2006}
{Bundy}, K., {Ellis}, R.~S., {Conselice}, C.~J., {et~al.} 2006, \apj, 651, 120

\bibitem[{{Cappelluti} {et~al.}(2013){Cappelluti}, {Kashlinsky}, {Arendt},
  {Comastri}, {Fazio}, {Finoguenov}, {Hasinger}, {Mather}, {Miyaji}, \&
  {Moseley}}]{Cappelluti2012}
{Cappelluti}, N., {Kashlinsky}, A., {Arendt}, R.~G., {et~al.} 2013, \apj, 769,
  68

\bibitem[{{Conroy} \& {Wechsler}(2009)}]{Conroy2009}
{Conroy}, C. \& {Wechsler}, R.~H. 2009, \apj, 696, 620

\bibitem[{{Cooray} \& {Sheth}(2002)}]{cooray2002}
{Cooray}, A. \& {Sheth}, R. 2002, \physrep, 372, 1

\bibitem[{{Cowie} {et~al.}(1996){Cowie}, {Songaila}, {Hu}, \&
  {Cohen}}]{Cowie1996}
{Cowie}, L.~L., {Songaila}, A., {Hu}, E.~M., \& {Cohen}, J.~G. 1996, \aj, 112,
  839

\bibitem[{{Croton} {et~al.}(2006){Croton}, {Springel}, {White}, {De Lucia},
  {Frenk}, {Gao}, {Jenkins}, {Kauffmann}, {Navarro}, \& {Yoshida}}]{Croton2006}
{Croton}, D.~J., {Springel}, V., {White}, S.~D.~M., {et~al.} 2006, \mnras, 365,
  11

\bibitem[{{de Putter} {et~al.}(2012){de Putter}, {Mena}, {Giusarma}, {Ho},
  {Cuesta}, {Seo}, {Ross}, {White}, {Bizyaev}, {Brewington}, {Kirkby},
  {Malanushenko}, {Malanushenko}, {Oravetz}, {Pan}, {Percival}, {Ross},
  {Schneider}, {Shelden}, {Simmons}, \& {Snedden}}]{dePutter2012}
{de Putter}, R., {Mena}, O., {Giusarma}, E., {et~al.} 2012, \apj, 761, 12

\bibitem[{{Dekel} \& {Birnboim}(2006)}]{Dekel2006}
{Dekel}, A. \& {Birnboim}, Y. 2006, \mnras, 368, 2

\bibitem[{{Dole} {et~al.}(2006){Dole}, {Lagache}, {Puget}, {Caputi},
  {Fern\`andez-Conde}, {Le Floc'h}, {Papovich}, {P\'erez-Gonz\`alez}, {Rieke},
  \& {Blaylock}}]{Dole2006}
{Dole}, H., {Lagache}, G., {Puget}, J.~L., {et~al.} 2006, \aap, 451, 417

\bibitem[{{Dunkley} {et~al.}(2011){Dunkley}, {Hlozek}, {Sievers}, {Acquaviva},
  {Ade}, {Aguirre}, {Amiri}, {Appel}, {Brown}, {Burger}, {Chervenak}, {Das},
  {Devlin}, {Dicker}, {Bertrand Doriese}, {D{\"u}nner}, {Essinger-Hileman},
  {Fisher}, {Fowler}, {Hajian}, {Halpern}, {Hasselfield},
  {Hern{\'a}ndez-Monteagudo}, {Hilton}, {Hilton}, {Hincks}, {Huffenberger},
  {Hughes}, {Hughes}, {Infante}, {Irwin}, {Juin}, {Kaul}, {Klein}, {Kosowsky},
  {Lau}, {Limon}, {Lin}, {Lupton}, {Marriage}, {Marsden}, {Mauskopf},
  {Menanteau}, {Moodley}, {Moseley}, {Netterfield}, {Niemack}, {Nolta}, {Page},
  {Parker}, {Partridge}, {Reid}, {Sehgal}, {Sherwin}, {Spergel}, {Staggs},
  {Swetz}, {Switzer}, {Thornton}, {Trac}, {Tucker}, {Warne}, {Wollack}, \&
  {Zhao}}]{Dunkley2011}
{Dunkley}, J., {Hlozek}, R., {Sievers}, J., {et~al.} 2011, \apj, 739, 52

\bibitem[{{Eisenstein} {et~al.}(2011){Eisenstein}, {Weinberg}, {Agol},
  {Aihara}, {Allende Prieto}, {Anderson}, {Arns}, {Aubourg}, {Bailey},
  {Balbinot}, \& et~al.}]{Eisenstein2011}
{Eisenstein}, D.~J., {Weinberg}, D.~H., {Agol}, E., {et~al.} 2011, \aj, 142, 72

\bibitem[{{Fall} \& {Efstathiou}(1980)}]{Fall1980}
{Fall}, S.~M. \& {Efstathiou}, G. 1980, \mnras, 193, 189

\bibitem[{{Fixsen} {et~al.}(1998){Fixsen}, {Dwek}, {Mather}, {Bennett}, \&
  {Shafer}}]{Fixsen1998}
{Fixsen}, D.~J., {Dwek}, E., {Mather}, J.~C., {Bennett}, C.~L., \& {Shafer},
  R.~A. 1998, \apj, 508, 123

\bibitem[{{Giannantonio} {et~al.}(2012){Giannantonio}, {Crittenden}, {Nichol},
  \& {Ross}}]{giannantonio2012}
{Giannantonio}, T., {Crittenden}, R., {Nichol}, R., \& {Ross}, A.~J. 2012,
  \mnras, 426, 2581

\bibitem[{{Giannantonio} {et~al.}(2008){Giannantonio}, {Scranton},
  {Crittenden}, {Nichol}, {Boughn}, {Myers}, \& {Richards}}]{Giannantonio2008}
{Giannantonio}, T., {Scranton}, R., {Crittenden}, R.~G., {et~al.} 2008, \prd,
  77, 123520

\bibitem[{{G{\'o}rski} {et~al.}(2005){G{\'o}rski}, {Hivon}, {Banday},
  {Wandelt}, {Hansen}, {Reinecke}, \& {Bartelmann}}]{Gorski2005}
{G{\'o}rski}, K.~M., {Hivon}, E., {Banday}, A.~J., {et~al.} 2005, \apj, 622,
  759

\bibitem[{{Guo} {et~al.}(2010){Guo}, {White}, {Li}, \&
  {Boylan-Kolchin}}]{Guo2010}
{Guo}, Q., {White}, S., {Li}, C., \& {Boylan-Kolchin}, M. 2010, \mnras, 404,
  1111

\bibitem[{{Hall} {et~al.}(2010){Hall}, {Keisler}, {Knox}, {Reichardt}, {Ade},
  {Aird}, {Benson}, {Bleem}, {Carlstrom}, {Chang}, {Cho}, {Crawford}, {Crites},
  {de Haan}, {Dobbs}, {George}, {Halverson}, {Holder}, {Holzapfel}, {Hrubes},
  {Joy}, {Lee}, {Leitch}, {Lueker}, {McMahon}, {Mehl}, {Meyer}, {Mohr},
  {Montroy}, {Padin}, {Plagge}, {Pryke}, {Ruhl}, {Schaffer}, {Shaw},
  {Shirokoff}, {Spieler}, {Stalder}, {Staniszewski}, {Stark}, {Switzer},
  {Vanderlinde}, {Vieira}, {Williamson}, \& {Zahn}}]{Hall2010}
{Hall}, N.~R., {Keisler}, R., {Knox}, L., {et~al.} 2010, \apj, 718, 632

\bibitem[{{Hauser} {et~al.}(1998){Hauser}, {Arendt}, {Kelsall}, {Dwek},
  {Reach}, {Silverberg}, {Moseley}, {Pei}, {Lubin}, {Mather}, {Shafer},
  {Smoot}, {Weiss}, {Wilkinson}, \& {Wright}}]{Hauser1998}
{Hauser}, M.~G., {Arendt}, R.~G., {Kelsall}, T., {et~al.} 1998, \apj, 508, 25

\bibitem[{{Hivon} {et~al.}(2002){Hivon}, {G{\'o}rski}, {Netterfield}, {Crill},
  {Prunet}, \& {Hansen}}]{Hivon2002}
{Hivon}, E., {G{\'o}rski}, K.~M., {Netterfield}, C.~B., {et~al.} 2002, \apj,
  567, 2

\bibitem[{{Ho} {et~al.}(2012){Ho}, {Cuesta}, {Seo}, {de Putter}, {Ross},
  {White}, {Padmanabhan}, {Saito}, {Schlegel}, {Schlafly}, {Seljak},
  {Hern{\'a}ndez-Monteagudo}, {S{\'a}nchez}, {Percival}, {Blanton}, {Skibba},
  {Schneider}, {Reid}, {Mena}, {Viel}, {Eisenstein}, {Prada}, {Weaver},
  {Bahcall}, {Bizyaev}, {Brewinton}, {Brinkman}, {Nicolaci da Costa}, {Gott},
  {Malanushenko}, {Malanushenko}, {Nichol}, {Oravetz}, {Pan},
  {Palanque-Delabrouille}, {Ross}, {Simmons}, {de Simoni}, {Snedden}, \&
  {Yeche}}]{Ho2012}
{Ho}, S., {Cuesta}, A., {Seo}, H.-J., {et~al.} 2012, \apj, 761, 14

\bibitem[{{Kashlinsky} {et~al.}(2012){Kashlinsky}, {Arendt}, {Ashby}, {Fazio},
  {Mather}, \& {Moseley}}]{Kash2012b}
{Kashlinsky}, A., {Arendt}, R.~G., {Ashby}, M.~L.~N., {et~al.} 2012, \apj, 753,
  63

\bibitem[{{Kashlinsky} {et~al.}(2005){Kashlinsky}, {Arendt}, {Mather}, \&
  {Moseley}}]{Kash2005}
{Kashlinsky}, A., {Arendt}, R.~G., {Mather}, J., \& {Moseley}, S.~H. 2005,
  \nat, 438, 45

\bibitem[{{Kashlinsky} \& {Odenwald}(2000)}]{Kash2000}
{Kashlinsky}, A. \& {Odenwald}, S. 2000, \apj, 528, 74

\bibitem[{{Kashlinsky} {et~al.}(2002){Kashlinsky}, {Odenwald}, {Mather},
  {Skrutskie}, \& {Cutri}}]{Kash2002}
{Kashlinsky}, A., {Odenwald}, S., {Mather}, J., {Skrutskie}, M.~F., \& {Cutri},
  R.~M. 2002, \apjl, 579, L53

\bibitem[{{Kere{\v s}} {et~al.}(2005){Kere{\v s}}, {Katz}, {Weinberg}, \&
  {Dav{\'e}}}]{Keres2005}
{Kere{\v s}}, D., {Katz}, N., {Weinberg}, D.~H., \& {Dav{\'e}}, R. 2005,
  \mnras, 363, 2

\bibitem[{{Lagache} {et~al.}(1999){Lagache}, {Abergel}, {Boulanger},
  {D{\'e}sert}, \& {Puget}}]{Lagache1999}
{Lagache}, G., {Abergel}, A., {Boulanger}, F., {D{\'e}sert}, F.~X., \& {Puget},
  J.-L. 1999, \aap, 344, 322

\bibitem[{{Lagache} {et~al.}(2007){Lagache}, {Bavouzet}, {Fernandez-Conde},
  {Ponthieu}, {Rodet}, {Dole}, {Miville-Desch{\^e}nes}, \&
  {Puget}}]{Lagache2007}
{Lagache}, G., {Bavouzet}, N., {Fernandez-Conde}, N., {et~al.} 2007, \apjl,
  665, L89

\bibitem[{{Lagache} {et~al.}(2000){Lagache}, {Haffner}, {Reynolds}, \&
  {Tufte}}]{Lagache2000}
{Lagache}, G., {Haffner}, L.~M., {Reynolds}, R.~J., \& {Tufte}, S.~L. 2000,
  \aap, 354, 247

\bibitem[{{Lagache} \& {Puget}(2000)}]{Lagache2000b}
{Lagache}, G. \& {Puget}, J.~L. 2000, \aap, 355, 17

\bibitem[{{Lewis} \& {Bridle}(2002)}]{lewis2002}
{Lewis}, A. \& {Bridle}, S. 2002, \prd, 66, 103511

\bibitem[{{Limber}(1953)}]{Limber1953}
{Limber}, D.~N. 1953, \apj, 117, 134

\bibitem[{{Matsuhara} {et~al.}(2000){Matsuhara}, {Kawara}, {Sato}, {Taniguchi},
  {Okuda}, {Matsumoto}, {Sofue}, {Wakamatsu}, {Cowie}, {Joseph}, \&
  {Sanders}}]{Matsuhara2000}
{Matsuhara}, H., {Kawara}, K., {Sato}, Y., {et~al.} 2000, \aap, 361, 407

\bibitem[{{Matsumoto} {et~al.}(2005){Matsumoto}, {Matsuura}, {Murakami},
  {Tanaka}, {Freund}, {Lim}, {Cohen}, {Kawada}, \& {Noda}}]{Matsumoto2005}
{Matsumoto}, T., {Matsuura}, S., {Murakami}, H., {et~al.} 2005, \apj, 626, 31

\bibitem[{{M{\'e}nard} {et~al.}(2013){M{\'e}nard}, {Scranton}, {Schmidt},
  {Morrison}, {Jeong}, {Budavari}, \& {Rahman}}]{menard2013}
{M{\'e}nard}, B., {Scranton}, R., {Schmidt}, S., {et~al.} 2013, ArXiv e-prints

\bibitem[{{Meny} {et~al.}(2007){Meny}, {Gromov}, {Boudet}, {Bernard},
  {Paradis}, \& {Nayral}}]{meny2007}
{Meny}, C., {Gromov}, V., {Boudet}, N., {et~al.} 2007, \aap, 468, 171

\bibitem[{{Miville-Desch{\^e}nes} \& {Lagache}(2005)}]{mamd2005}
{Miville-Desch{\^e}nes}, M.-A. \& {Lagache}, G. 2005, \apjs, 157, 302

\bibitem[{{Miville-Desch{\^e}nes} {et~al.}(2002){Miville-Desch{\^e}nes},
  {Lagache}, \& {Puget}}]{mivlag2002}
{Miville-Desch{\^e}nes}, M.-A., {Lagache}, G., \& {Puget}, J.-L. 2002, \aap,
  393, 749

\bibitem[{{Moster} {et~al.}(2010){Moster}, {Somerville}, {Maulbetsch}, {van den
  Bosch}, {Macci{\`o}}, {Naab}, \& {Oser}}]{Moster2010}
{Moster}, B.~P., {Somerville}, R.~S., {Maulbetsch}, C., {et~al.} 2010, \apj,
  710, 903

\bibitem[{{Navarro} {et~al.}(1997){Navarro}, {Frenk}, \&
  {White}}]{1997ApJ...490..493N}
{Navarro}, J.~F., {Frenk}, C.~S., \& {White}, S.~D.~M. 1997, \apj, 490, 493

\bibitem[{{P{\^a}ris} {et~al.}(2013){P{\^a}ris}, {Petitjean}, {Aubourg},
  {Ross}, {Myers}, {Streblyanska}, {Bailey}, {Hall}, {Strauss}, {Anderson},
  {Bizyaev}, {Borde}, {Brinkmann}, {Bovy}, {Brandt}, {Brewington},
  {Brownstein}, {Cook}, {Ebelke}, {Fan}, {Filiz Ak}, {Finley}, {Font-Ribera},
  {Ge}, {Hamann}, {Ho}, {Jiang}, {Kinemuchi}, {Malanushenko}, {Malanushenko},
  {Marchante}, {McGreer}, {McMahon}, {Miralda-Escud{\'e}}, {Muna},
  {Noterdaeme}, {Oravetz}, {Palanque-Delabrouille}, {Pan}, {Perez-Fournon},
  {Pieri}, {Riffel}, {Schlegel}, {Schneider}, {Simmons}, {Viel}, {Weaver},
  {Wood-Vasey}, {Y{\`e}che}, \& {York}}]{Paris2013}
{P{\^a}ris}, I., {Petitjean}, P., {Aubourg}, {\'E}., {et~al.} 2013, ArXiv
  e-prints

\bibitem[{{Planck Collaboration I}(2013)}]{planck2013-p01}
{Planck Collaboration I}. 2013, Submitted to \aap, [arXiv:astro-ph/1303.5062]

\bibitem[{{Planck Collaboration VI}(2013)}]{planck2013-p03}
{Planck Collaboration VI}. 2013, Submitted to \aap, [arXiv:astro-ph/1303.5067]

\bibitem[{{Planck Collaboration VII}(2013)}]{planck2013-p03c}
{Planck Collaboration VII}. 2013, Submitted to \aap, [arXiv:astro-ph/1303.5068]

\bibitem[{{Planck Collaboration VIII}(2013)}]{planck2013-p03f}
{Planck Collaboration VIII}. 2013, Submitted to \aap,
  [arXiv:astro-ph/1303.5069]

\bibitem[{{Planck Collaboration XVI}(2013)}]{planck2013-p11}
{Planck Collaboration XVI}. 2013, Submitted to \aap, [arXiv:astro-ph/1303.5076]

\bibitem[{{Planck Collaboration XVIII}(2011)}]{planck2011-6.6}
{Planck Collaboration XVIII}. 2011, \aap, 536, A18

\bibitem[{{Planck Collaboration XVIII}(2013)}]{planck2013-p13}
{Planck Collaboration XVIII}. 2013, Submitted to \aap,
  [arXiv:astro-ph/1303.5078]

\bibitem[{{Planck Collaboration XXVIII}(2013)}]{planck2013-p05}
{Planck Collaboration XXVIII}. 2013, Submitted to \aap,
  [arXiv:astro-ph/1303.5088]

\bibitem[{{Planck Collaboration XXX}(2013)}]{planck2013-pip56}
{Planck Collaboration XXX}. 2013, Submitted to \aap, [arXiv:astro-ph/1309.0382]

\bibitem[{{Puget} {et~al.}(1996){Puget}, {Abergel}, {Bernard}, {Boulanger},
  {Burton}, {Desert}, \& {Hartmann}}]{Puget1996}
{Puget}, J.-L., {Abergel}, A., {Bernard}, J.-P., {et~al.} 1996, \aap, 308, L5

\bibitem[{{Rees} \& {Ostriker}(1977)}]{Rees1977}
{Rees}, M.~J. \& {Ostriker}, J.~P. 1977, \mnras, 179, 541

\bibitem[{{Reichardt} {et~al.}(2012){Reichardt}, {Shaw}, {Zahn}, {Aird},
  {Benson}, {Bleem}, {Carlstrom}, {Chang}, {Cho}, {Crawford}, {Crites}, {de
  Haan}, {Dobbs}, {Dudley}, {George}, {Halverson}, {Holder}, {Holzapfel},
  {Hoover}, {Hou}, {Hrubes}, {Joy}, {Keisler}, {Knox}, {Lee}, {Leitch},
  {Lueker}, {Luong-Van}, {McMahon}, {Mehl}, {Meyer}, {Millea}, {Mohr},
  {Montroy}, {Natoli}, {Padin}, {Plagge}, {Pryke}, {Ruhl}, {Schaffer},
  {Shirokoff}, {Spieler}, {Staniszewski}, {Stark}, {Story}, {van Engelen},
  {Vanderlinde}, {Vieira}, \& {Williamson}}]{Reichardt2012}
{Reichardt}, C.~L., {Shaw}, L., {Zahn}, O., {et~al.} 2012, \apj, 755, 70

\bibitem[{{Ross} {et~al.}(2011){Ross}, {Ho}, {Cuesta}, {Tojeiro}, {Percival},
  {Wake}, {Masters}, {Nichol}, {Myers}, {de Simoni}, {Seo},
  {Hern{\'a}ndez-Monteagudo}, {Crittenden}, {Blanton}, {Brinkmann}, {da Costa},
  {Guo}, {Kazin}, {Maia}, {Maraston}, {Padmanabhan}, {Prada}, {Ramos},
  {Sanchez}, {Schlafly}, {Schlegel}, {Schneider}, {Skibba}, {Thomas}, {Weaver},
  {White}, \& {Zehavi}}]{Ross2011}
{Ross}, A.~J., {Ho}, S., {Cuesta}, A.~J., {et~al.} 2011, \mnras, 417, 1350

\bibitem[{{Sawangwit} {et~al.}(2010){Sawangwit}, {Shanks}, {Cannon}, {Croom},
  {Ross}, \& {Wake}}]{sawangwit2010}
{Sawangwit}, U., {Shanks}, T., {Cannon}, R.~D., {et~al.} 2010, \mnras, 402,
  2228

\bibitem[{{Shang} {et~al.}(2012){Shang}, {Haiman}, {Knox}, \& {Oh}}]{shang2012}
{Shang}, C., {Haiman}, Z., {Knox}, L., \& {Oh}, S.~P. 2012, \mnras, 421, 2832

\bibitem[{{Silk}(1977)}]{Silk1977}
{Silk}, J. 1977, \apj, 211, 638

\bibitem[{{Silk}(2003)}]{Silk2003}
{Silk}, J. 2003, \mnras, 343, 249

\bibitem[{{Smith} {et~al.}(2003){Smith}, {Peacock}, {Jenkins}, {White},
  {Frenk}, {Pearce}, {Thomas}, {Efstathiou}, \& {Couchman}}]{Smith2003}
{Smith}, R.~E., {Peacock}, J.~A., {Jenkins}, A., {et~al.} 2003, \mnras, 341,
  1311

\bibitem[{{Tinker} {et~al.}(2008){Tinker}, {Kravtsov}, {Klypin}, {Abazajian},
  {Warren}, {Yepes}, {Gottl{\"o}ber}, \& {Holz}}]{tinker2008}
{Tinker}, J., {Kravtsov}, A.~V., {Klypin}, A., {et~al.} 2008, \apj, 688, 709

\bibitem[{{Tinker} {et~al.}(2010){Tinker}, {Robertson}, {Kravtsov}, {Klypin},
  {Warren}, {Yepes}, \& {Gottl{\"o}ber}}]{tinker2010}
{Tinker}, J.~L., {Robertson}, B.~E., {Kravtsov}, A.~V., {et~al.} 2010, \apj,
  724, 878

\bibitem[{{Viero} {et~al.}(2009){Viero}, {Ade}, {Bock}, {Chapin}, {Devlin},
  {Griffin}, {Gundersen}, {Halpern}, {Hargrave}, {Hughes}, {Klein},
  {MacTavish}, {Marsden}, {Martin}, {Mauskopf}, {Moncelsi}, {Negrello},
  {Netterfield}, {Olmi}, {Pascale}, {Patanchon}, {Rex}, {Scott}, {Semisch},
  {Thomas}, {Truch}, {Tucker}, {Tucker}, \& {Wiebe}}]{Viero2009}
{Viero}, M.~P., {Ade}, P.~A.~R., {Bock}, J.~J., {et~al.} 2009, \apj, 707, 1766

\bibitem[{{Viero} {et~al.}(2013){Viero}, {Wang}, {Zemcov}, {Addison},
  {Amblard}, {Arumugam}, {Aussel}, {B{\'e}thermin}, {Bock}, {Boselli}, {Buat},
  {Burgarella}, {Casey}, {Clements}, {Conley}, {Conversi}, {Cooray}, {De
  Zotti}, {Dowell}, {Farrah}, {Franceschini}, {Glenn}, {Griffin},
  {Hatziminaoglou}, {Heinis}, {Ibar}, {Ivison}, {Lagache}, {Levenson},
  {Marchetti}, {Marsden}, {Nguyen}, {O'Halloran}, {Oliver}, {Omont}, {Page},
  {Papageorgiou}, {Pearson}, {P{\'e}rez-Fournon}, {Pohlen}, {Rigopoulou},
  {Roseboom}, {Rowan-Robinson}, {Schulz}, {Scott}, {Seymour}, {Shupe}, {Smith},
  {Symeonidis}, {Vaccari}, {Valtchanov}, {Vieira}, {Wardlow}, \&
  {Xu}}]{viero2012}
{Viero}, M.~P., {Wang}, L., {Zemcov}, M., {et~al.} 2013, \apj, 772, 77

\bibitem[{{Wang} {et~al.}(2014){Wang}, {Viero}, {Ross}, {Asboth}, {Bethermin},
  {Bock}, {Clements}, {Conley}, {Cooray}, {Farrah}, {Hajian}, {Han}, {Lagache},
  {Marsden}, {Myers}, {Norberg}, {Oliver}, {Page}, {Symeonidis}, {Schulz},
  {Wang}, \& {Zemcov}}]{Wang2014}
{Wang}, L., {Viero}, M., {Ross}, N.~P., {et~al.} 2014, ArXiv e-prints

\bibitem[{{White} {et~al.}(2011){White}, {Blanton}, {Bolton}, {Schlegel},
  {Tinker}, {Berlind}, {da Costa}, {Kazin}, {Lin}, {Maia}, {McBride},
  {Padmanabhan}, {Parejko}, {Percival}, {Prada}, {Ramos}, {Sheldon}, {de
  Simoni}, {Skibba}, {Thomas}, {Wake}, {Zehavi}, {Zheng}, {Nichol},
  {Schneider}, {Strauss}, {Weaver}, \& {Weinberg}}]{White2011}
{White}, M., {Blanton}, M., {Bolton}, A., {et~al.} 2011, \apj, 728, 126

\bibitem[{{White} \& {Rees}(1978)}]{White1978}
{White}, S.~D.~M. \& {Rees}, M.~J. 1978, \mnras, 183, 341

\bibitem[{{Wright} {et~al.}(2010){Wright}, {Eisenhardt}, {Mainzer}, {Ressler},
  {Cutri}, {Jarrett}, {Kirkpatrick}, {Padgett}, {McMillan}, {Skrutskie},
  {Stanford}, {Cohen}, {Walker}, {Mather}, {Leisawitz}, {Gautier}, {McLean},
  {Benford}, {Lonsdale}, {Blain}, {Mendez}, {Irace}, {Duval}, {Liu}, {Royer},
  {Heinrichsen}, {Howard}, {Shannon}, {Kendall}, {Walsh}, {Larsen}, {Cardon},
  {Schick}, {Schwalm}, {Abid}, {Fabinsky}, {Naes}, \& {Tsai}}]{wright2010}
{Wright}, E.~L., {Eisenhardt}, P.~R.~M., {Mainzer}, A.~K., {et~al.} 2010, \aj,
  140, 1868

\bibitem[{{Xia} {et~al.}(2012){Xia}, {Negrello}, {Lapi}, {De Zotti}, {Danese},
  \& {Viel}}]{xia2012}
{Xia}, J.-Q., {Negrello}, M., {Lapi}, A., {et~al.} 2012, \mnras, 422, 1324

\end{thebibliography}

\end{document}